\documentclass[12pt]{article}

\usepackage{graphicx}
\usepackage{float}
\usepackage{cite}
\usepackage{amsfonts}
\usepackage{amssymb}
\usepackage{xcolor}

\usepackage{amsmath}
\usepackage[margin=1.4in]{geometry}

\def\gtwid{\mathrel{\raise.3ex\hbox{$>$\kern-.75em\lower1ex\hbox{$\sim$}}}}
\def\ltwid{\mathrel{\raise.3ex\hbox{$<$\kern-.75em\lower1ex\hbox{$\sim$}}}}
\def\square{\kern1pt\vbox{\hrule height 1.2pt\hbox{\vrule width 1.2pt\hskip 3pt
   \vbox{\vskip 6pt}\hskip 3pt\vrule width 0.6pt}\hrule height 0.6pt}\kern1pt}

\def\gtwid{\mathrel{\raise.3ex\hbox{$>$\kern-.75em\lower1ex\hbox{$\sim$}}}}
\def\ltwid{\mathrel{\raise.3ex\hbox{$<$\kern-.75em\lower1ex\hbox{$\sim$}}}}
\def\square{\kern1pt\vbox{\hrule height 1.2pt\hbox{\vrule width 1.2pt\hskip 3pt
   \vbox{\vskip 6pt}\hskip 3pt\vrule width 0.6pt}\hrule height 0.6pt}\kern1pt}

\begin{document}

\begin{titlepage}

\begin{center}
{\bf K-Essence Induced by Derivative Couplings of the Inflaton}
\end{center}

\vskip 2cm

\begin{center}
Y. S. Hung$^{\dagger}$, S. P. Miao$^{\ddagger}$
\end{center}

\vskip 1cm

\begin{center}
\it{Department of Physics, National Cheng Kung University, \\
No. 1 University Road, Tainan City 70101, TAIWAN}
\end{center}

\vspace{2cm}

\begin{center}
ABSTRACT
\end{center}
We consider two models which couple derivatives of the inflaton to
ordinary matter, both to fermions and to scalars. Such couplings induce changes
to the inflaton kinetic energy, analogous to the cosmological Coleman-Weinberg
potentials which come from nonderivative couplings. Our purpose is to investigate
whether these quantum-induced K-Essence models can provide efficient reheating
without affecting the observational constraints on primordial inflation.
Our numerical studies show that it is difficult to preserve both properties.

\begin{flushleft}
\end{flushleft}

\vskip 4cm

\begin{flushleft}
$^{\dagger}$ email: L26104034@gs.ncku.edu.tw\\
$^{\ddagger}$ e-mail: spmiao5@mail.ncku.edu.tw
\end{flushleft}

\end{titlepage}

\section{Introduction}
Single scalar inflation is the simplest model consistent with current data,
\begin{equation}
\mathcal{L} = \frac{R \sqrt{-g}}{16 \pi G} - \frac12 \partial_{\mu} \varphi
\partial_{\nu} \varphi g^{\mu\nu} \sqrt{-g} - V(\varphi) \sqrt{-g} \; .
\label{GRMCS}
\end{equation}
Given a desired expansion history $a(t)$, one can construct a scalar potential
and an initial condition which will support it \cite{Tsamis:1997rk,Saini:1999ba,
Capozziello:2005mj}. This is important because the observational constraints on
inflation can be phrased in terms of the expansion history $a(t)$ and its
derivatives,
\begin{equation}
ds^2 = -dt^2 + a^2(t) d\vec{x} \!\cdot\!d\vec{x} \qquad \Longrightarrow \qquad
H(t) \equiv \frac{\dot{a}}{a} \quad , \quad \epsilon(t) \equiv -\frac{\dot{H}}{H^2}
\; . \label{geometry}
\end{equation}
These constraints are, first, that the number of e-foldings from the start of
inflation at $t_i$ to its end (when $\epsilon(t_e) = 1$) should be large enough
to explain the Horizon Problem,
\begin{equation}
N \equiv \ln\Bigl[ \frac{a(t_e)}{a(t_i)}\Bigr] \gtwid 60 \; . \label{duration}
\end{equation}
Additional constraints come from the slow roll approximations for the scalar
and tensor power spectra,
\begin{equation}
\Delta^2_{\mathcal{R}}(k) \simeq \frac{G H^2(t_k)}{\pi \epsilon(t_k)}
\qquad , \qquad \Delta^2_{h}(k) \simeq \frac{16}{\pi} G H^2(t_k) \; ,
\label{spectra}
\end{equation}
where $t_k$ is the time of first horizon crossing at which $k = a(t_k) H(t_k)$.
The observed scalar perturbations experience first crossing over a period of
about ten e-foldings, starting about 50 e-foldings before the end of inflation.
Near the beginning of this period the value of $\Delta^2_{\mathcal{R}}(k)$ must
be about $2 \times 10^{-9}$, and the scalar spectral index $n_s$ must obey
\cite{Planck:2018vyg},
\begin{equation}
1- n_s \equiv -\frac{\partial \ln(\Delta^2_{\mathcal{R}})}{\partial \ln(k)}
\simeq 2 \epsilon + \frac{\dot{\epsilon}}{H \epsilon} \simeq 0.035 \; .
\label{scalarspec}
\end{equation}
Finally, the non-detection of primordial tensors implies \cite{Planck:2018vyg},
\begin{equation}
r \equiv \frac{\Delta^2_h}{\Delta^2_{\mathcal{R}}} \simeq 16 \epsilon \ltwid
0.036 \; . \label{tensortoscalar}
\end{equation}

The observational constraints (\ref{duration}-\ref{tensortoscalar}) result in
very flat potentials, whose minimum is infinitesimally close to zero, and with
initial conditions which seem unnaturally fine-tuned to some people. However, our
concern here is the {\it additional} constraints which arise from coupling the
inflaton to ordinary matter in order to facilitate reheating. The simplest
couplings involve the undifferentiated inflaton, for example, a Yukawa coupling
$\varphi \overline{\Psi} \Psi$ to fermions. What happens then is that vacuum
fluctuations of ordinary matter induce Coleman-Weinberg potentials that are
neither Planck-suppressed nor limited to local functionals of metric which
could be completely eliminated by local counterterms \cite{MW1}. Because they
are not Planck-suppressed, these cosmological Coleman-Weinberg potentials
typically cause dramatic changes in the inflationary expansion history which
endangers the observational constraints (\ref{duration}-\ref{tensortoscalar}).

Although cosmological Coleman-Weinberg potentials cannot be completely
eliminated by allowed counterterms, two partial subtraction schemes are
possible:
\begin{itemize}
{\item {\bf Hubble subtraction} in which a local function of the inflaton is
used to null quantum effects at the onset of inflation \cite{LMW}; and}
{\item {\bf Ricci subtraction} in which a local function of the inflaton and
the Ricci scalar are used to null quantum effects for $\epsilon = 0$ \cite{MPW}}.
\end{itemize}
\noindent Neither technique gives good results. Employing Hubble subtraction
\cite{LMW} shows that, with a moderate coupling constant, inflation never ends
for fermionic couplings of effective potential, and ends too soon for vector
boson couplings. Making the coupling constants very small results in acceptable
inflation at the price of inefficient reheating. Ricci subtraction gives even
worse results \cite{MPW}. With this scheme neither model experiences more than
a single e-folding of inflation, no matter how small the coupling constant. This
is because Ricci subtraction introduces an extra, unsuppressed degree of
freedom which makes a fatal change in the first Friedmann equation.

Because nonderivative couplings are so problematic we have decided here to
explore the consequences of derivative couplings.\footnote{This idea was
suggested by the un-named referee of \cite{LMW} to whom we are grateful.}
Because the inflaton oscillates during reheating, its derivative should be
just about as effective as the undifferentiated field at communicating kinetic
energy to ordinary matter. Of course a derivative coupling will not induce a
Coleman-Weinberg potential; it will instead generate a nonlinear function of
the inflaton kinetic energy, which is a kind of the quantum-induced K-Essence
model \cite{Armendariz-Picon:1999hyi,Garriga:1999vw,Babichev:2007dw}.
Our goal is to check whether the resulting models, with Hubble subtraction,
can resolve the inconsistency between efficiency of the reheating and the
observational constraints on inflation.

This paper consists of six sections, of which the first is nearly done.
In section 2 we consider a model with derivative couplings of the inflaton
to fermions, and work out the induced K-Essence. Section 3 does the same for
couplings to scalars. The two modified Friedmann equations and the scalar
evolution equation are derived in section 4. Section 5 investigates the
effective kinetic energy induced by fermions and scalars. Our conclusions
are presented in section 6.

\section{Model with Derivative Coupling of the Inflaton to Fermions}

The inflaton could be derivative-coupled to a masssless fermion $\Psi(x)$,
\begin{equation}
\mathcal{L}_{\rm fermion} = \overline{\Psi} \gamma^b e^{\mu}_{~ b}
\Bigl( \partial_{\mu} \!+\! \frac{i}2 A_{\mu cd} J^{cd} \Bigr) \Psi
\sqrt{-g} + \frac{1}{2 m_c^3}\partial_{\mu} \varphi
\partial_{\nu} \varphi g^{\mu\nu}\overline{\Psi} \Psi \sqrt{-g} \; .
\label{fermionL}
\end{equation}
Here $e^{\mu}_{~b}(x)$ is the vierbein field with $g^{\mu\nu}(x)
= e^{\mu}_{~b}(x) e^{\nu}_{~c}(x) \eta^{bc}$ and $A_{\mu cd}(x) =
e^{\nu}_{~c} [e_{\nu d , \mu} - \Gamma^{\rho}_{~\mu\nu} e_{\rho d}]$
is the spin connection. The symbol $\gamma^b_{ij}$ represents the $4
\times 4$ gamma matrices which obey $\{ \gamma^b,\gamma^c\} = -2
\eta^{bc} I$, and $J^{cd} \equiv \frac{i}{4} [\gamma^c,\gamma^d]$
are the Lorentz representation matrices for Dirac fermions. The final term
is the coupling between derivative inflaton and fermions
and $\frac{1}{m_c^3}$ is strength of the dimensionful coupling.

Taking the functional derivative of (\ref{GRMCS}) $+$ (\ref{fermionL}),
and then replacing $\overline{\Psi} \Psi$ with the coincident fermion
propagator gives the effective field equation,
\begin{eqnarray}
\partial_{\mu}\Big[\sqrt{-g}g^{\mu\nu}\partial_{\nu}\varphi \Big]
-V'(\varphi)\sqrt{-g}+\frac{1}{m_c^3}\partial_{\mu}
\Big[\sqrt{-g}g^{\mu\nu}\partial_{\nu}\varphi
\!\times\!\textrm{Tr}\,\{iS[M](x,x)\}\Big]\!=0
\,.\label{eomf0}
\end{eqnarray}
The sign flip in the final term is due to the definition of the fermion
propagator, $i[_i S_j](x;x')\!\equiv
\!\langle\Psi_i(x)\overline{\Psi}_j(x')\rangle$.
To compute the final term of the effective equation,
we employ the coincidence limit of the massive fermion propagator on
de Sitter background ($\epsilon = 0$) \cite{CandelasRaine,MW2},
\begin{eqnarray}
iS[M](x;x)\!=\!\frac{H^{D-2}}{(4\pi)^{\frac{D}{2}}}
\Gamma\Big(1\!-\!\frac{D}{2}\Big)
\frac{\Gamma(\frac{D}{2}\!+\!i\frac{M}{H})\Gamma(\frac{D}{2}\!-\!i\frac{M}{H})}
{\Gamma(1\!+\!i\frac{M}{H})\Gamma(1\!-\!i\frac{M}{H})}\!\times\! MI\,.
\label{Fcoincidence}
\end{eqnarray}
Here $M \equiv \frac{-1}{2 m_c^3}\partial_{\mu}\varphi\partial_{\nu}\varphi
g^{\mu\nu}$ is the fermion mass and $I$ stands for the identity matrix.
Expanding expression (\ref{Fcoincidence})
around $D\!=\!4$ and substituting to (\ref{eomf0}), the unregulated limit of
the quantum-induced term can be obtained,
\begin{eqnarray}
\frac{\partial_{\mu}}{m_c^3}
\left\{\begin{matrix}
\!\!\sqrt{-g}g^{\mu\nu}\partial_{\nu}\varphi\!\times\!\!
\left[\begin{matrix}
\hskip -5.4cm\frac{H^{D-2}}{(4\pi)^{\frac{D}{2}}}
\Gamma\Big(1\!-\!\frac{D}{2}\Big)\Big[4M\!+4\frac{M^3}{H^2}\Big]+\\
\frac{H^2}{16\pi^2}\Bigg\{8M\!+\!\Big[4M\!+\!4\frac{M^3}{H^2}\Big]
\Big[\psi(1\!+\!i\frac{M}{H})\!+\!\psi(1\!-\!i\frac{M}{H})\Big]
\Bigg\}\!+\mathcal{O}(D\!-\!4)
\end{matrix}\right]
\end{matrix}\right\}.\label{unrenormal0}
\end{eqnarray}
Note that $\psi$ in the second line is not a fermion field but rather the digamma
function $\psi(z) \equiv \frac{d}{dz} \ln[\Gamma(z)]$.

One can see that two counterterms are needed to renormalize
the first line of (\ref{unrenormal0}),
\begin{eqnarray}
&&\mathcal{L}_{\rm ct}\!=\!-\frac14\delta z_1
\Big(\partial_{\mu}\varphi\partial_{\nu}\varphi g^{\mu\nu}\Big)^2\!\sqrt{-g}
-\frac18 \delta z_2
(\partial_{\mu}\varphi\partial_{\nu}\varphi g^{\mu\nu})^4\sqrt{-g}\,,\nonumber\\
&&\frac{\delta S_{\rm ct}}{\delta\varphi}\Longrightarrow
\delta z_1 \partial_{\mu}\Big[\sqrt{-g}g^{\mu\nu}\partial_{\nu}\varphi
(\partial_{\rho}\varphi\partial_{\sigma}\varphi g^{\rho\sigma})\Big]
\!+\!\delta z_2 \partial_{\mu}\Big[\sqrt{-g}g^{\mu\nu}\partial_{\nu}\varphi
(\partial_{\rho}\varphi\partial_{\sigma}\varphi g^{\rho\sigma})^3\Big]\,.
\nonumber\\
\label{f2ct}
\end{eqnarray}
We make the following choice of $\delta z_1$ and $\delta z_2$,
\begin{eqnarray}
&&\delta z_1\!=\!\frac{H^{D-2}}{(4\pi)^{\frac{D}{2}}}
\Gamma\Big(1\!-\!\frac{D}{2}\Big)\frac{2}{m_c^6}
+\frac{H^2}{4\pi^2}\frac{(1\!-\!\gamma)}{m_c^6}\,,\nonumber\\
&&\delta z_2\!=\!\frac{H^{D-2}}{(4\pi)^{\frac{D}{2}}}
\frac{\Gamma(1\!-\!\frac{D}{2})}{2m_c^{12}H^2}
+\frac{H^2}{16\pi^2}
\frac{[-\gamma\!+\!\zeta(3)]}{m_c^{12}H^2}\,,\label{f2ctcoeff}
\end{eqnarray}
in order to absorb the divergences and to eliminate the two lowest-order terms
in the small field expansion of (\ref{unrenormal0}). After combining
(\ref{f2ctcoeff}), (\ref{f2ct}) with (\ref{unrenormal0}), the re-normalized
result is,
\begin{eqnarray}
\partial_{\mu}
\left\{\begin{matrix}
\!\sqrt{-g}g^{\mu\nu}\partial_{\nu}\varphi\!\times\!\!\frac{H^4}{16\pi^2}\!\!
\left[\begin{matrix}
\hskip -3cm \frac{8\gamma}{m_c^3 H}
\Big(\frac{M}{H}\Big)
\!+\!\frac{8\gamma-8\zeta(3)}{m_c^{3} H}
\Big(\frac{M}{H}\Big)^{\!\!3}
\\
\!+\Big[\frac{4}{m_c^3 H}
\Big(\!\frac{M}{H}\!\Big)
\!+\!\frac{4}{m_c^{3} H}
\Big(\!\frac{M}{H}\!\Big)^{\!\!3}
\Big]\Big[\psi(1\!+\!i\frac{M}{H})\!+\!\psi(1\!-\!i\frac{M}{H})\Big]
\end{matrix}\right]
\end{matrix}\right\}.\label{renormal0}
\end{eqnarray}

This quantum contribution in the field equation can be regarded as a kind
of quantum-induced K-Essence. Suppose that there exists a function of the
kinetic term in the action
$ \int d^D x\, \Delta K(-\frac12 \partial_{\rho}\varphi\partial_{\sigma}\varphi
g^{\rho\sigma})\sqrt{-g}$,
then the contribution to the effective field equation gives,
\begin{eqnarray}
\partial_{\mu}\Big[\sqrt{-g}g^{\mu\nu}\partial_{\nu}\varphi\times
\Delta K'(-\frac12 \partial_{\rho}\varphi\partial_{\sigma}\varphi g^{\rho\sigma})\Big]\,.
\label{Kprime}
\end{eqnarray}
We can immediately identify (\ref{renormal0}) as
$\Delta K'(-\frac12 \partial_{\rho}\varphi\partial_{\sigma}\varphi g^{\rho\sigma})$.
By integrating (\ref{renormal0}) back with the argument, the quantum-induced
kinetic term is,
\begin{eqnarray}
&&\Delta K_{\rm f}(z) \!= \!\frac{H^4}{8\pi^2}\Bigg\{2\gamma z^2 +[\gamma\!-\!\zeta(3)]z^4
\!+2\!\!\int_0^z\!\! dx\, [x\!+\!x^2]\Big[\psi(1\!+\!ix)+\psi(1\!-\!ix)\Big]
\Bigg\}\,, \nonumber \\
&& \textrm{where z is defined as}\;
z\!\equiv\!-\frac{\partial_{\rho}\varphi\partial_{\sigma}\varphi g^{\rho\sigma}}
{2m_c^3 H}\!=\!\frac{M}{H}\,.\label{fermion}
\end{eqnarray}
To get the small field expansion we substitute,
\begin{equation}
\vert z \vert \ll 1 \Longrightarrow \psi(1 \!+\! z) = - \gamma -
\sum_{k = 1}^{\infty} \zeta(k \!+\! 1) (-z)^k \; . \qquad \label{smallpsi}
\end{equation}
The resulting expansion is,
\begin{eqnarray}
&&\Delta K_{\rm f}(z) = \frac{H^4}{4 \pi^2} \sum_{n=2}^{\infty}
\frac{(-1)^n}{n \!+\! 1} \Bigl[ \zeta(2n \!-\! 1) - \zeta(2n \!+\! 1)\Bigr]
z^{2n+2} \; , \qquad \\
&&\hskip 1.5cm =  \frac{H^4}{8\pi^2} \Biggl\{ \frac23 \Bigl[\zeta(3) \!-\! \zeta(5)\Bigr]
z^6 -\frac1{2} \Bigl[\zeta(5) \!-\! \zeta(7)\Bigr] z^8
+ \mathcal{O}(z^{10})\Biggr\} .\label{Ksmallf}
\qquad
\end{eqnarray}
By substituting the large argument expansion for the digamma function,
\begin{eqnarray}
|z|\gg1 \Longrightarrow \psi(z)\!=\!\ln(z)-\frac{1}{2z}-\frac{1}{12z^2}
+\frac{1}{120z^4}-\frac{1}{256z^6}+\mathcal{O}(\frac{1}{z^8})\,,
\label{largepsi}
\end{eqnarray}
the large field expansion can be obtained,
\begin{eqnarray}
&&\Delta K_{\rm f}(z) = \frac{H^4}{8 \pi^2} \Biggl\{
\frac12 z^4 \ln(z^2 \!+\! 1) - \Bigl( \zeta(3) \!+\! \frac14 \!-\! \gamma\Bigr)
z^4 + z^2 \ln(z^2 \!+\! 1) \nonumber \\
& & \hspace{4.8cm} - \Bigl( \frac43 \!-\! 2\gamma\Bigr) z^2 + \frac{11}{60}
\ln(z^2 \!+\! 1) + O(z^0) \Biggr\} . \qquad \label{Klargef}
\end{eqnarray}

\section{The Model with Derivative Coupling of the Inflaton to Scalars}

In this section, we begin with a derivation of the quantum-induced K-Essence
model due to scalars with arbitrary non-minimal coupling. We then
present two special cases. One is a conformal coupling and the other
is the minimally-coupled one.

\subsection{A General Derivation with Non-Minimal Coupling}

Derivatives of the inflaton $\varphi(x)$ might couple to another scalar $\Phi(x)$, which
need not be minimally coupled to gravity\footnote{The $+$ sign is chosen for stability
because $M^2 \equiv -\frac1{m^2_c} \partial_{\mu} \varphi \partial_{\nu} \varphi
g^{\mu\nu}$ is positive for a time-dependent inflaton.},
\begin{eqnarray}
\mathcal{L}_{\phi}\!=\!-\frac12\partial_{\mu}\Phi\partial_{\nu}\Phi g^{\mu\nu}\!\sqrt{-g}
-\frac{1}{12}(1\!+\!\Delta\xi)R\Phi^2\sqrt{-g}+\frac{1}{2m_c^2}
\partial_{\mu}\varphi\partial_{\nu}\varphi g^{\mu\nu}\Phi^2\sqrt{-g}\,,\label{phi}
\end{eqnarray}
where $\frac1{m_c^2}$ is a dimensionful coupling strength. Note that $\Delta\xi\!=\!0$
corresponds to a conformally coupled scalar. The mass term of the scalar $\Phi(x)$
can be identified from (\ref{phi}),
\begin{eqnarray}
M^2_{\Phi}\!=\!-\frac{1}{m_c^2}
\partial_{\mu}\varphi\partial_{\nu}\varphi g^{\mu\nu}\,.\label{mphi}
\end{eqnarray}
Taking a functional derivative of (\ref{GRMCS}) $+$ (\ref{phi}), and replacing $\Phi^2$
with the coincidence limit of the $\Phi$ propagator gives the effective field equation,
\begin{eqnarray}
\partial_{\mu}\Big[\sqrt{-g}g^{\mu\nu}\partial_{\nu}\varphi \Big]
-V'(\varphi)\sqrt{-g}-\frac{1}{m_c^2}\partial_{\mu}
\Big[\sqrt{-g}g^{\mu\nu}\partial_{\nu}\varphi
\!\times\! i\Delta[\xi,M^2_{\Phi}](x;x)\Big]=0\,.\label{eom0}
\end{eqnarray}
On de Sitter background ($\epsilon = 0$) the coincidence limit of the scalar
propagator is \cite{MW1,MTW},
\begin{eqnarray}
i\Delta[\xi,M^2_{\Phi}](x;x)\!=\!
\frac{H^{D-2}}{(4\pi)^{\frac{D}{2}}}\Gamma\Big(1\!-\!\frac{D}{2}\Big)
\frac{\Gamma(\frac{D-1}{2}\!+\!\nu)\Gamma(\frac{D-1}{2}\!-\!\nu)}
{\Gamma(\frac12\!+\!\nu)\Gamma(\frac12\!-\!\nu)}\,,\label{propxx}
\end{eqnarray}
where $\nu^2$ is,
\begin{eqnarray}
\nu^2\!=\!\Big(\frac{D-1}{2}\Big)^2-D(D-1)\xi-\frac{M^2_{\Phi}}{H^2}\,\,\,,\,\,\,
\xi\equiv\frac16(1+\Delta\xi)\,.\label{nuxi}
\end{eqnarray}
After substituting (\ref{propxx}) in (\ref{eom0}), expanding around $D\!=\!4$
and segregating finite parts from divergences, we see that two counterterms are
needed to renormalize the primitive contribution,
\begin{eqnarray}
&&\mathcal{L}_{\rm ct}\!=\!-\frac12\delta Z \partial_{\mu}\varphi\partial_{\nu}\varphi
g^{\mu\nu}\!\sqrt{-g}-\frac14 \delta Z_1
(\partial_{\mu}\varphi\partial_{\nu}\varphi g^{\mu\nu})^2\sqrt{-g}\,,\nonumber\\
&&\frac{\delta S_{\rm ct}}{\delta\varphi}\Longrightarrow\,\,\,
\delta Z \partial_{\mu}\Big[\sqrt{-g}g^{\mu\nu}\partial_{\nu}\varphi\Big]
+\delta Z_1 \partial_{\mu}\Big[\sqrt{-g}g^{\mu\nu}\partial_{\nu}\varphi
(\partial_{\rho}\varphi\partial_{\sigma}\varphi g^{\rho\sigma})\Big]\,.\label{2ct}
\end{eqnarray}
Here the finite parts of $\delta Z$ and $\delta Z_1$ are chosen to cancel
the $\partial_{\rho}\varphi\partial_{\sigma}\varphi g^{\rho\sigma}$ and
$(\partial_{\rho}\varphi\partial_{\sigma}\varphi g^{\rho\sigma})^2$ terms
in the small field expansion\footnote{The expression of the finite parts
works for $\Delta\xi\geqq 0$.},
\begin{eqnarray}
&& \hskip -0.6cm \delta Z\!=\!\frac{1}{m_c^2}\Bigg\{
\frac{H^{D-2}}{(4\pi)^{\frac{D}{2}}}\Gamma\Big(1\!-\!\frac{D}{2}\Big)\!
\Big(\! 2\Delta\xi\Big)
+\frac{H^2}{16\pi^2}\Big(\frac13+\frac73\Delta\xi\!+\!
2\Delta\xi[\psi(\nu_{+})\!+\!\psi(\nu_{-})]\Big)\Bigg\}\,,\nonumber\\
&&\hskip -0.6cm\delta Z_1\!=\!\frac{-1}{m_c^2}\Bigg\{\frac{H^{D-2}}{(4\pi)^{\frac{D}{2}}}
\frac{\Gamma(1\!-\!\frac{D}{2})}{m_c^2H^2}+\frac{H^2}{16\pi^2}\Bigg(
\frac{[\psi(\nu_{+})\!+\!\psi(\nu_{-})]}{m_c^2H^2}
\!-\!\frac{2\Delta\xi[\psi'(\nu_{+})\!-\!\psi'(\nu_{-})]}
{m_c^2H^2\sqrt{1-8\Delta\xi}}\Bigg)\Bigg\}\,.\nonumber \\
\label{2ctcoeff}
\end{eqnarray}
Note that $\psi(z) \equiv \frac{d}{dz} \ln[\Gamma(z)]$ is the digamma function
and that we define $\nu_{\pm}\equiv \frac12\pm\frac12\sqrt{1-8\Delta\xi}$.
The renormalized result can be expressed in terms of the dimensionless
quantity $y\equiv \frac{-\partial_{\mu}\varphi\partial_{\nu}\varphi g^{\mu\nu}}
{2m_c^2H^2}$,
\begin{eqnarray}
&&\hskip -0.7cm\Delta K(y)\!=\!\frac{H^4}{16\pi^2}\Bigg\{\Big[2\Delta\xi y\!+\!y^2\Big]
\Big[\psi(\nu_{+})\!+\!\psi(\nu_{-})\Big]-
\frac{2\Delta\xi y^2[\psi'(\nu_{+})\!-\!\psi'(\nu_{-})]}{\sqrt{1-8\Delta\xi}}
\nonumber\\
&&\hskip -1cm -\,2\!\!\int_0^{y}\!\!dx (\Delta\xi\!+\!x)
\Bigg[\psi\Big(\frac12\!+\!\frac12\sqrt{1\!-\!8(\Delta\xi\!+\!x)}\Big)
+ \psi\Big(\frac12\!-\!\frac12\sqrt{1\!-\!8(\Delta\xi\!+\!x)}\Big)\Bigg]
\Bigg\}.\label{K}
\end{eqnarray}
We Taylor expand the digamma function in (\ref{K}) to get the
the small field expansion,
\begin{eqnarray}
&&\Delta K(y)\!=\!\frac{H^4}{16\pi^2}\Bigg\{\Bigg[
\frac{(1\!-\!6\Delta\xi)[\psi'(\nu_{+})\!-\!\psi'(\nu_{-})]}{(1\!-\!8\Delta\xi)^{\frac32}}
-\frac{\Delta\xi[\psi''(\nu_{+})\!+\!\psi''(\nu_{-})]}{(1\!-\!8\Delta\xi)}\Bigg]
\Big(\frac{4y^3}{3}\Big)
\nonumber\\
&&\hspace{2cm}+\Bigg[\frac{2(1\!-\!4\Delta\xi)[\psi'(\nu_{+})\!-\!\psi'(\nu_{-})]}
{(1\!-\!8\Delta\xi)^{\frac52}}-\frac{(1\!-\!4\Delta\xi)[\psi''(\nu_{+})\!+\!\psi''(\nu_{-})]}
{2(1\!-\!8\Delta\xi)^{2}}\nonumber\\
&&\hspace{5cm}+\frac{\frac23\Delta\xi[\psi'''(\nu_{+})\!-\!\psi'''(\nu_{-})]}
{(1\!-\!8\Delta\xi)^{\frac32}}\Bigg]y^4
+\mathcal{O}(y^{5})\Bigg\}\,.\label{Ksmall}
\end{eqnarray}
The large field expansion comes from using the asymptotic expansion (\ref{largepsi}),
\begin{eqnarray}
&&\hskip -1cm \Delta K(y)\!=\!\frac{H^4}{16\pi^2}
\Bigg\{\!\!-\!y^2\ln\!\Big(\!2y\!+\!2\Delta\xi\Big)\!+\!
\Bigg[\frac12\!+\!\psi(\nu_{+})\!+\!\psi(\nu_{-})
-\frac{2\Delta\xi[\psi'(\nu_{+})\!-\!\psi'(\nu_{-})]}
{\sqrt{1\!-\!8\Delta\xi}}\Bigg]y^2\nonumber\\
&&\hskip 3cm -2\Delta\xi y\ln\!\Big(\!2y\!+\!2\Delta\xi\Big)
\!+\!\Bigg[\frac13\!+\!\Delta\xi
\!+\!2\Delta\xi\Big[\psi(\nu_{+})\!+\!\psi(\nu_{-})\Big]\Bigg]y\nonumber\\
&&\hskip 5.5cm+\Big(\frac{1}{30}\!-\!\Delta\xi^2 \Big)\ln\Big(2y\!+\!2\Delta\xi\Big)
+\mathcal{O}(y^0)\Bigg\}\,.\label{Klarge}
\end{eqnarray}

\subsection{Conformal Coupling}

In this subsection, we specialize to the case in which derivatives of the
inflaton are coupled to a massless, conformally coupled scalar.
By expanding expression (\ref{K}) for small $\Delta\xi$, and carefully dealing with
singular contributions from the digamma functions and its derivatives such as,
\begin{eqnarray}
&&\psi(\nu_{-})\simeq\frac{-1}{2\Delta\xi}+1-\gamma+\cdots\,,\\
&&\psi'(\nu_{-})\simeq\frac{1}{4\Delta\xi^2}(1-4\Delta\xi-4\Delta\xi^2)
+\frac{\pi^2}{6}+\cdots\,,
\end{eqnarray}
the quantum-induced term is seen to be,
\begin{eqnarray}
\Delta K_{\rm con}(y)\!=\! \frac{H^4}{16\pi^2}
\Bigg\{\!\!\!-y\!+\!(1\!-\!2\gamma)y^2
\!-\!2\!\!\int_0^y\!\!\! dx \,x\Big[
\psi\Big(\frac12\!+\!\frac12\sqrt{1\!-\!8x}\!\Big)\!+\!
\psi\Big(\frac12\!-\!\frac12\sqrt{1\!-\!8x}\!\Big) \Big]\!\Bigg\}.
\label{scalarCon}
\end{eqnarray}
Employing the same techniques as the proceeding section gives
the small field and large field expansions,
\begin{eqnarray}
&&\hskip -1cm \Delta K_{\rm conS}(y)\!=\!\frac{H^4}{16\pi^2}\Bigg\{
\!\!-\!\frac43 y^3\!-\!\Big[4+2\psi''(1)\Big]y^4\!+\!\mathcal{O}(y^5)
\Bigg\}\,,\\
&&\hskip -1cm\Delta K_{\rm conL}(y)\!=\!\frac{H^4}{16\pi^2}\Bigg\{\!\!
-y^2\ln(2y)\!+\!\Big(\frac32\!-\!2\gamma\Big)y^2-\frac23 y+\frac{1}{30}\ln(2y)
\!+\!\mathcal{O}(y^0)
\Bigg\}\,.\label{ConL}
\end{eqnarray}

\subsection{Minimal Coupling}

The other special case we consider is minimal coupling, for which
$\Delta\xi$ is taken to $-1$ in expression (\ref{phi}). We assume
that $M_{\phi}$ is small and positive because the massless limit of a
massive, minimally coupled propagator is not smooth \cite{Block,
BunchDavis,FolacciAllen}. Furthermore, in order to eliminate the two
leading contributions in the small field expansion, the finite parts
of (\ref{2ctcoeff}) need to be adjusted as follows,
\begin{eqnarray}
&& \hskip -0.6cm \delta \mathcal{Z}\!=\!\frac{1}{m_c^2}\Bigg\{
\frac{H^{D-2}}{(4\pi)^{\frac{D}{2}}}
\Big[\!\!-2\Gamma\Big(1\!-\!\frac{D}{2}\Big)\Big]\!
+\frac{H^2}{16\pi^2}\Big(4\gamma\!-\!\frac{29}{3}\Big)\Bigg\}\,,\nonumber\\
&&\hskip -0.6cm\delta \mathcal{Z}_1\!=\!
\frac{-1}{m_c^2}\Bigg\{\frac{H^{D-2}}{(4\pi)^{\frac{D}{2}}}
\frac{\Gamma(1\!-\!\frac{D}{2})}{m_c^2H^2}+\frac{H^2}{16\pi^2}\Bigg(
\frac{-2\gamma+\frac{25}{27}}{m_c^2H^2}\Bigg)\Bigg\}\,.
\label{Mini2ctcoeff}
\end{eqnarray}
A straightforward but lengthy computation gives the exact result,
\begin{eqnarray}
&&\Delta K_{\rm min}(y)\!=\! \frac{H^4}{16\pi^2}
\Bigg\{\!\!-2\gamma\!+\!\frac{182}{27}
\!+\!\Big(4\gamma\!-\!\frac{23}{3}\Big)y
\!+\!\Big(-2\gamma\!+\!\frac{25}{27}\Big)y^2 \nonumber\\
&&+2\!\!\int_1^y\!\!\! dx \,(1\!-\!x)\Bigg[
\psi\Bigg(\frac12\!+\!\frac32\sqrt{1\!-\!\frac89 x}\!\Bigg)\!+\!
\psi\Bigg(\frac12\!-\!\frac32\sqrt{1\!-\!\frac89 x}\!\Bigg) \Bigg]\Bigg\}.
\label{scalarminimal}
\end{eqnarray}
The small and large field expansions are,
\begin{eqnarray}
\hskip -1cm \Delta K_{\rm minS}(y)\!=\!\frac{H^4}{16\pi^2}
\Bigg\{\!\!-3\ln(y)\!+\!C\!+\!\Big[\frac{-92}{729}\!+\!\frac{8}{27}\psi''(1)\Big]y^3
\!-\frac{2}{81}\!\Big[\frac{44}{27}\!+\!5\psi''(1)\Big]y^4\!+\!\cdots
\!\Bigg\},\nonumber\\
\end{eqnarray}
\begin{eqnarray}
&&\hskip -0.1cm\Delta K_{\rm minL}(y)\!=\!\frac{H^4}{16\pi^2}\Bigg\{\!\!
-y^2\ln(2y)\!+\!\Big(\frac{77}{54}\!-\!2\gamma\Big)y^2
\!+\!2y\ln(2y)\nonumber\\
&&\hskip 5cm+\!\Big[4\gamma-\frac{25}{3}\Big] y\!-\!\frac{29}{30}\ln(2y)
\!+\!\mathcal{O}(y^0)
\Bigg\}\,.\label{minL}
\end{eqnarray}
Here $C\!=\!\frac{14}{81}[\frac{26}{27}-\psi''(1)]$
is an integration constant.

\section{The Modified Friedmann Equations}

Our results (\ref{fermion}), (\ref{K}), (\ref{scalarCon}) and (\ref{scalarminimal})
were all derived on de Sitter background because that is the only case for which
the necessary propagators are known. However, for realistic inflation the first
slow roll parameter is nonzero and the Hubble parameter changes with time. Numerical
studies of cosmological Coleman-Weinberg potentials have shown that it is reasonable
to simply replace the constant de Sitter $H_{\rm dS}$ with the evolving $H(t)$, and
ignore $\epsilon(t)$ \cite{Kyriazis:2019xgj,Sivasankaran:2020dzp,Katuwal:2021kry}.
That is what we shall do for quantum-induced K-Essence models,
\begin{eqnarray}
&& \mathcal{L} = \frac{R \sqrt{-g}}{16 \pi G}\!+\!K(\textsc{k},H)\sqrt{-g}
- V(\varphi) \sqrt{-g} \; ,\nonumber\\
&& K(\textsc{k},H)\equiv \textsc{k}\!+\!\Delta K(\textsc{k},H)\,,\,\,\,\,
\textsc{k}=- \frac12 \partial_{\mu} \varphi\partial_{\nu} \varphi g^{\mu\nu}\,.
\label{GRK}
\end{eqnarray}
The purpose of this section is to work out how the two Friedmann equations and the
scalar evolution equation change. The section closes by converting these equations
into a dimensionless form conducive to numerical work.

If the effective action is known for a general metric $g_{\mu\nu}(x)$ the modified
Friedmann equations can be obtained by taking the functional derivative with respect
to it and then specializing to the cosmological background (\ref{geometry}).
However, what we have is (an approximation for) the effective action already
specialized to (\ref{geometry}). Because this depends only upon the single
gravitational dynamical variable $a(t)$, varying can give at most one of the two
Friedmann equations. The theorem of Palais \cite{Palais,Torre} guarantees us that
the single equation so obtained is at least correct. We first show that this is
the second Friedmann equation, then we use conservation to reconstruct the first
Friedmann equation.

In our homogeneous, isotropic and spatially flat geometry (\ref{geometry}) the
Lagrangian (\ref{GRK}) becomes,
\begin{eqnarray}
L \!=\!-\frac{6a^3H^2}{16\pi G}+a^3 K(\textsc{k}, H)-a^3V(\varphi)\,.
\end{eqnarray}
The gravitational dynamical variable is $a(t)$ and the associated
Euler-Lagrange equation is,
\begin{eqnarray}
-2\dot{H}\!-\!3H^2\!=\!8\pi G\Big[K(\textsc{k},H)\!-\!V(\varphi)
\!-\!H\frac{\partial K(\textsc{k},H)}{\partial H}-\frac13\frac{d}{dt}
\frac{\partial K(\textsc{k},H)}{\partial H}\Big]\,.\label{ij}
\end{eqnarray}
This is the $ij$ Einstein equation, sometimes known as the second Friedmann
equation. Varying (\ref{GRK}) with respect to $\varphi(t)$ gives the scalar
evolution equation,
\begin{eqnarray}
\ddot{\varphi}\Big(2\textsc{k}\frac{\partial^2 K}{\partial \textsc{k}^2}
\!+\!\frac{\partial K}{\partial \textsc{k}}\Big)\!+\!\dot{\varphi}
\Big(\dot{H}\frac{\partial^2 K}{\partial\textsc{k}\partial H}
\!+\!3H\frac{\partial K}{\partial \textsc{k}}\Big)
\!+\!\frac{\partial V}{\partial\varphi}=0\,.\label{scalar0}
\end{eqnarray}
Missing is the first Friedmann equation, which is the $00$ Einstein
equation. We can recover it by noting that the two Friedmann equations
are related to the scalar evolution equation through conservation,
\begin{eqnarray}
\hskip -0.2cm\frac{d}{dt}\Big[00\, \rm{equation}\Big]
\!+\!3H\Big[(00\!+\!ij)\;\rm{equations}\Big]\!=\!8\pi G\dot{\varphi}
\Big[\rm{scalar\; evolution\; equation}\Big].\label{relation}
\end{eqnarray}
From relation (\ref{relation}) we infer the first Friedmann equation,
\begin{eqnarray}
3H^2\!=\!8\pi G \Big[2\textsc{k}\frac{\partial K}{\partial \textsc{k}}
\!-\!K(\textsc{k},H)\!+\!H\frac{\partial K}{\partial H}\!+\!V(\varphi)\Big]\,.
\label{00}
\end{eqnarray}

Because the scale of temporal variation changes dramatically over the course
of inflation, and because the dependent variables $\varphi(t)$ and $H(t)$ are
dimensionful, it is convenient to convert to dimensionless variables. We
first change the co-moving time $t$ to the number of the inflationary
e-foldings since the beginning of inflation,
\begin{equation}
n \equiv \ln\Bigl( \frac{a(t)}{a(t_i)}\Bigr) \quad \Longrightarrow \quad
\frac{d}{dt} = H \frac{d}{d n} \quad , \quad \frac{d^2}{d t^2} = H^2
\Bigl[ \frac{d^2}{d n^2} - \epsilon \frac{d}{d n}\Bigr] \; .
\end{equation}
It is also useful to make the various other quantities dimensionless,
\begin{eqnarray}
&&\phi(n) \!\equiv\! \sqrt{8 \pi G} \,\varphi(t) \,\, , \,\, \chi(n)
\!\equiv\! \sqrt{8 \pi G} \,H(t) \,\,,\,\,
\kappa(n)\!\equiv\!(8\pi G)^2\textsc{k}(t)
\!=\!\frac12\chi^2\phi'^2\,,\nonumber\\
&& k_c \!\equiv\! \sqrt{8 \pi G} \, m_c \;\; , \;\;
\mathcal{K}(\kappa,\chi)\!\equiv\!(8\pi G)^2\,K(\textsc{k},H)
\,\,,\,\,U(\phi)\!\equiv\!(8\pi G)^2\,V(\varphi)\,.
\label{dimensionless}
\end{eqnarray}
With these changes, the two modified Friedman equations (\ref{ij}) and (\ref{00})
take the forms,
\begin{eqnarray}
&&3\chi^2\!=2\kappa\frac{\partial \mathcal{K}}{\partial \kappa}
\!-\!\mathcal{K}(\kappa,\chi)\!+\!\chi\frac{\partial \mathcal{K}}{\partial \chi}
\!+\!U(\phi)\,,\label{F1}\\
&&-2\chi\chi'\!-\!3\chi^2\!=\! \mathcal{K}(\kappa,\chi)\!-\!U(\phi)
\!-\!\chi\frac{\partial \mathcal{K}}{\partial \chi}
\!-\!\frac13\chi\frac{d}{dn}
\frac{\partial \mathcal{K}(\kappa,\chi)}{\partial \chi}\,.\label{F2}
\end{eqnarray}
And the scalar evolution equation becomes,
\begin{eqnarray}
\phi''\Big(2\kappa\frac{\partial^2 \mathcal{K}}{\partial \kappa^2}
\!+\!\frac{\partial \mathcal{K}}{\partial \kappa}\Big)\!+\!\phi'
\Big[\!-2\kappa\epsilon\frac{\partial^2\mathcal{K}}{\partial\kappa^2}
\!-\!\epsilon\chi\frac{\partial^2 \mathcal{K}}{\partial\kappa\partial \chi}
\!+\!(3\!-\!\epsilon)\frac{\partial \mathcal{K}}{\partial \kappa}\Big]
\!+\!\frac{1}{\chi^2}\frac{\partial U}{\partial\phi}=0\,,\label{scalar1}
\end{eqnarray}
where the first slow roll parameter is expressed as,
\begin{eqnarray}
\epsilon(n) \!\equiv\! -\frac{\chi'}{\chi}\! =\!
\frac{ \frac12 {\phi'}^2\,\frac{\partial\mathcal{K}}{\partial\kappa}
\!-\!\frac16 \chi\phi''\phi'\,
\frac{\partial^2 \mathcal{K}}{\partial \kappa \partial \chi}}
{1 \!-\! \frac16 \Big[\frac{\partial^2 \mathcal{K}}{\partial \chi^2}
\!+\!\chi\phi'^2\frac{\partial^2\mathcal{K}}{\partial\kappa\partial\chi}\Big]}
\!=\!\frac{\frac{{\phi'}^2}{2}\frac{\partial\mathcal{K}}{\partial\kappa}
\Big[\frac{\partial\mathcal{K}}{\partial\kappa}
\!+\!2\kappa\frac{\partial^2\mathcal{K}}{\partial\kappa^2}
\!+\!\chi\frac{\partial^2\mathcal{K}}{\partial\chi\partial\kappa}\Big]
\!+\!\frac16\frac{\phi'}{\chi}
\frac{\partial^2\mathcal{K}}{\partial\chi\partial\kappa}
\frac{\partial U}{\partial\phi}}
{\Big[\frac{\partial\mathcal{K}}{\partial\kappa}
\!+\!2\kappa\frac{\partial^2\mathcal{K}}{\partial\kappa^2}\Big]
\Big[1\!-\!\frac16\frac{\partial^2\mathcal{K}}{\partial\kappa^2}\Big]
\!+\!\frac{\kappa}{3}
\Big(\frac{\partial^2\mathcal{K}}{\partial\chi\partial\kappa}\Big)^{\!2}}
\;.\nonumber\\ \label{epsilon}
\end{eqnarray}
Here we have used equations (\ref{F1}) and (\ref{F2}) in the first equality
and eliminated $\phi''$ to obtain the final expression using
the scalar evolution equation (\ref{scalar1}).

\section{The Fate of the $m^2\varphi^2$ Model\label{Fate}}

Because the Hubble parameter $H(t)$ is not a local functional of the metric,
we cannot completely subtract the quantum-induced K-Essence
$\Delta K(\textsc{k}(t),H(t))$ in expression (\ref{GRK}) using permissible
counterterms. What we could do instead is to subtract $\Delta K(\textsc{k}(t),
H_i)$, which would restore the classical model at the initial time, but would
lead to quantum corrections as evolution carries $H(t)$ away from $H(t_i)$.
This procedure is known as {\bf Hubble subtraction}. The purpose of this section
is to investigate the effects of quantum-induced K-Essence models with Hubble
subtraction in the context of the classical $m^2\varphi^2$ model. This model
has the virtue of simplicity, even though it is not consistent with the current
upper bound (\ref{tensortoscalar}) on the tensor-to-scalar ratio $r$
\cite{Planck2018, Improvedr}. We begin by describing classical evolution of the
$m^2\varphi^2$ model, then consider the effects of fermionic-induced K-Essence
(\ref{fermion}), and finally close by discussing the corrections due to
conformal scalars (\ref{scalarCon}) and minimally-coupled scalars
(\ref{scalarminimal}).

The dimensionless expressions of the two classical Friedman equations and
the scalar evolution equation are,
\begin{eqnarray}
&&3\chi^2\!=\!\frac12\chi^2\phi'^2\!+\!U(\phi)\;\;\;,\;\;\;
(2\epsilon\!-\!3)\chi^2\!=\!\frac12\chi^2\phi'^2\!-\!U(\phi)\,,\\
&&\phi''\!+\!(3\!-\!\epsilon)\phi'\!+\!\frac{1}{\chi^2}
\frac{\partial U(\phi)}{\partial\phi}=0\,,
\end{eqnarray}
where $\chi$, $\phi$ are defined in (\ref{dimensionless})
and $U(\phi)$ can be expressed as,
\begin{eqnarray}
U(\phi)\!=\!\frac12 k^2\phi^2\,\,,\,\,
k^2\!\equiv\!8\pi G m^2\,.
\end{eqnarray}
Under the slow roll approximation, the time evolutions of several useful
quantities can be found,
\begin{eqnarray}
&&\phi(n)\!\simeq\!\sqrt{\phi^2_0\!-\!4n}\,\,\,,\,\,\,\phi'(n)\!\simeq\!\frac{-2}{\phi(n)}
\,\,\,,\,\,\,\epsilon(n)\!\simeq\!\frac{2}{\phi(n)^2}\,\,\,,\,\,\,
\epsilon'(n)\!\simeq\!2\epsilon^2\,,\label{phiSlowRoll}\\
&&\Delta^2_{\mathcal{R}} \! \simeq \! \frac1{8\pi^2} \frac{\chi^2}{\epsilon}
\longrightarrow \frac1{8\pi^2} \frac{k^2}{3 \epsilon^2} \;\; ,\;\;
 1 - n_s \! \simeq \! 2 \epsilon + \frac{\epsilon'}{\epsilon} \longrightarrow
 4 \epsilon \; . \label{DRns}
\end{eqnarray}
The justification for the magnitude of the dimensionless mass $k$ is to reproduce
the scalar power spectrum amplitude $A_s$ and spectral index $n_s$ \cite{Planck2018},
\begin{equation}
k \simeq \pi (1 \!-\! n_s) \sqrt{\frac32 A_s} \simeq 6.13 \times 10^{-6} \; .
\label{kvalue}
\end{equation}
To make inflation last about $100$ e-foldings, the slow roll approximation
suggests initial conditions,
\begin{eqnarray}
\phi_{0}=20\;\;\;,\;\;\;\phi'_{0}=\frac{-1}{10}\;\;\;,\;\;\;
\chi_{0}\simeq\frac{1}{\sqrt{6}}\,k\phi_{0}\simeq 5\times10^{-5}
\,. \label{initials}
\end{eqnarray}
These should continue to apply in the quantum-corrected model as long as
the classical kinetic term $K_C\!\equiv\!\kappa$ dominates over
the quantum correction $K_Q\!\equiv\!\Delta\mathcal{K}$.
In principle, two initial conditions are enough to numerically simulate
the system, even with the quantum correction. However, because the first
Friedmann equation (\ref{F1}) is highly nonlinear in $\chi$, it is simpler
to evolve the quantum-corrected system using equations (\ref{scalar1}) and
(\ref{epsilon}). That is why we need an initial condition for $\chi_0$.

For the regime where quantum contributions $K_Q$ become comparable or
larger than the classical result $K_C$, one should fix {\it geometric}
initial conditions, making $\chi_0$ and $\epsilon_0$ equal their classical
values. From the first modified Friedmann equation (\ref{F1}), the potential
still dominates the right-hand side below some coupling strength \footnote{
For fermionic K-Essence, the kinetic contributions become comparable with
the potential contribution around $1/k^3_c \sim 1.564\!\times\!10^{10}$. }
so it leaves $\chi_0$ unchanged. Using the second expression of (\ref{epsilon}),
one can find $\epsilon_0$ as a function of $\phi'_0$ at the specific coupling
strength with fixed values of $\phi_0$ and $\chi_0$ and then solve for
$\phi'_0$ numerically by demanding $\epsilon_0(\phi'_0)\!=\!
[\epsilon_0]_{\rm classical} \!\sim\!5\!\times\!10^{-3}$. One should use
this more appropriate $\phi'_0$ in the quantum-dominated regime.

\subsection{Induced K-Essence Model due to Fermions}

If fermions quantum-correct the kinetic term and Hubble subtraction
is employed, its dimensionless form is,
\begin{equation}
\mathcal{K}(\kappa,\chi) = \kappa \!+\! \frac{\chi^4}{8 \pi^2}
f\Bigl(\frac{\kappa}{k^3_c \chi}\Bigr) \!-\! \frac{\chi^4(0)}{8\pi^2}
f\Bigl(\frac{\kappa}{k^3_c \chi(0)}\Bigr) \; , \label{fermiK}
\end{equation}
where $f(z)$ can be identified from (\ref{fermion}),
\begin{equation}
f(z) = 2 \gamma z^2 + \Bigl[ \gamma-\zeta(3)\Bigr] z^4 + 2 \int_{0}^{z}
\!\!\! dx \, (x \!+\! x^3) \Bigl[ \psi(1 \!+\! i x) + \psi(1 \!-\! i x)\Bigr]
\; . \label{fermif}
\end{equation}
At this point we digress to discuss the magnitude of the dimensionless
coupling strength $k_c \equiv \sqrt{8 \pi G} \, m_c$. Our goal was to make
the derivative coupling roughly comparable to the non-derivative coupling
during reheating,
\begin{eqnarray}
\frac{1}{2 m_c^3}\partial_{\mu} \varphi
\partial_{\nu} \varphi g^{\mu\nu}\overline{\Psi} \Psi \sqrt{-g}
\qquad \Longleftrightarrow \qquad
-\lambda \varphi \overline{\Psi} \Psi \sqrt{-g} \, .
\end{eqnarray}
Because the inflaton depends only on time we infer,
\begin{eqnarray}
\lambda\varphi\sim\frac{\dot{\varphi}^2}{2m_c^3}\,.
\end{eqnarray}
The scale of $\frac12 \dot{\varphi}^2$ can be approximated as
$\frac12 \omega^2\varphi^2$ and it is roughly equal to
$\frac12 m^2\varphi^2$. Therefore the dimensionless parameter $k_c$ relates
to the dimensionless Yukawa coupling as,
\begin{eqnarray}
\frac{1}{k_c^3}\sim\lambda\Big(\frac{2}{k^2\phi}\Big)
\sim (5.32\times10^{12})\lambda\,. \label{lambdadef}
\end{eqnarray}
Here we have used the relation (\ref{kvalue}) and $\phi\sim 10^{-2}$
that is roughly valid during reheating. Finally note that the cosmological
Coleman-Weinberg potential due to the fermion corrections has a similar
expression to (\ref{fermiK}) \cite{MW1,LMW,CandelasRaine},
\begin{eqnarray}
U(\phi,\chi) = \frac12 k^2 \phi^2 \!-\! \frac{\chi^4}{8 \pi^2} f\Bigl(
\frac{\lambda \phi}{\chi}\Bigr) + \frac{\chi^4(0)}{8\pi^2} f\Bigl(
\frac{\lambda \phi}{\chi(0)}\Bigr) \; , \label{fermiU}
\end{eqnarray}
where the form of the function $f$ is exactly the same as (\ref{fermif}).

\begin{figure}[H]
\centering
\includegraphics[width=11cm]{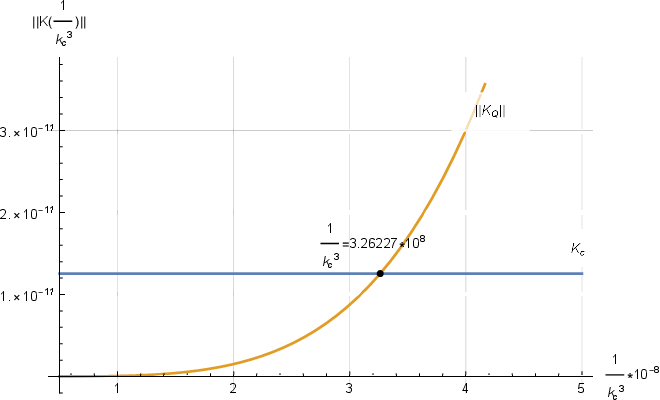}
\caption{\footnotesize Comparison of the initial ($n=0$) magnitudes of the
classical kinetic energy (in blue) with the quantum correction (in yellow)
for different values of the dimensionless coupling $1/k_c^3$, assuming
slow-roll initial conditions (\ref{initials}).}
\label{absKq+Kc-fermion}
\end{figure}

There is an interesting distinction between cosmological Coleman-Weinberg
potentials and quantum-induced K-Essence. Because the inflaton rolls down
its potential, cosmological Coleman-Weinberg potentials become smaller as
inflation progresses. However, the inflaton's kinetic energy does not show
a similar decline, meaning that quantum-induced K-Essence terms do not
typically fall off. Obviously, increasing $1/k_c^3$ makes the quantum
contribution larger. Figure \ref{absKq+Kc-fermion} shows that the quantum
correction begins dominating the initial kinetic energy around
$1/k^3_c\!\sim\!3.26\!\times\!10^8$, which corresponds to a small Yukawa
coupling $\lambda\!\sim\!6.13\!\times\!10^{-5}$.

\begin{figure}[H]
\includegraphics[width=6.5cm,height=4.6cm]{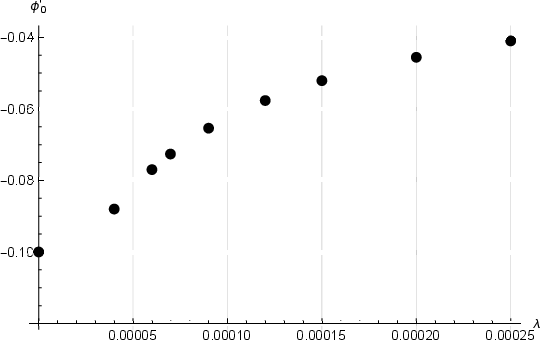}
\hspace{0.4cm}
\includegraphics[width=6.5cm,height=4.6cm]{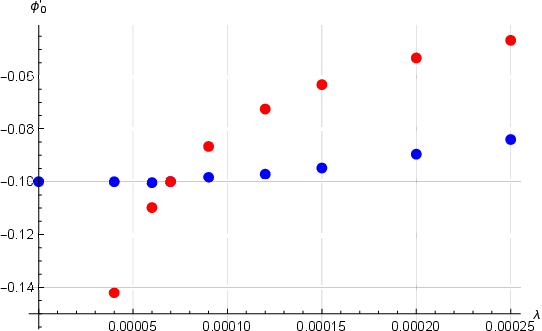}
\caption{\footnotesize The initial condition $\phi'_0$ in the quantum-dominated
regime versus the equivalent coupling constant $\lambda$ (\ref{lambdadef}).
The left-hand graph depicts a series of solutions for the un-subtracted model.
The two branches of the subtracted model are shown in the right-hand plot.}
\label{Branches-phi'_0}
\end{figure}

In spite of the quantum correction dominating the classical kinetic energy
it is still possible to stay close to the classical evolution provided the
slow-roll initial conditions (\ref{initials}) are abandoned for geometric
conditions as described
at the end of section \textbf{\ref{Fate}}. As long as the potential term
dominates over the rest of the kinetic contributions in equation (\ref{F1}),
choosing $\chi_0\!=\!5\!\times\!10^{-5}$ is still a reasonable approximation,
whereas $\phi'_0$ needs to be solved numerically by demanding that the initial
first slow roll parameter equals its classical value. Without Hubble
subtraction (the un-subtracted model), only one solution is found to give
$\epsilon(\phi'_0)\!=\!5\!\times\!10^{-3}$. With Hubble subtraction
(the subtracted model), the non-linearity of the function $\epsilon(\phi'_0)$
results in two solutions at each specific coupling strength. One should choose
the branch which is connected to the classical value $\phi'_0\!=\!-0.1$ for
small $1/k_c^3$. This is shown by the blue dots in Figure \ref{Branches-phi'_0}.

\begin{figure}[H]
\includegraphics[width=7.5cm,height=4.2cm]{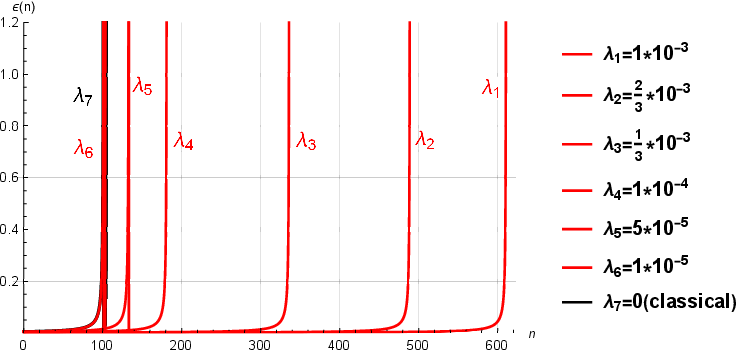}
\hspace{-0.2cm}
\includegraphics[width=7.5cm,height=4.2cm]{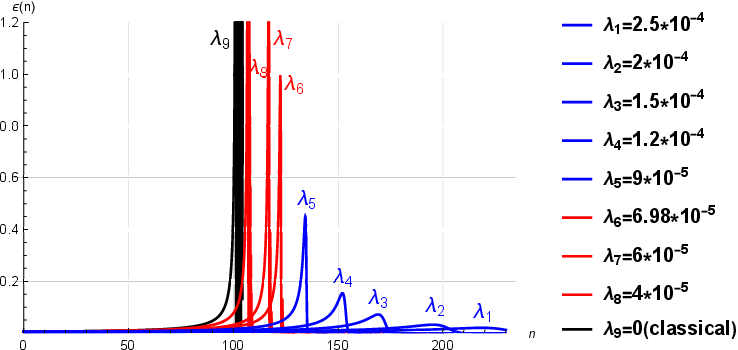}
\caption{\footnotesize The $\epsilon(n)$ versus $n$ for various $\lambda$'s.
The un-subtracted model is on the left-hand side while the subtracted one is plotted on
the right hand.}
\label{eps-n-variousLlambdas}
\end{figure}

We chose several coupling strengths within a range such that
$\chi_0\!\sim\!5\!\times\!10^{-5}$ is still valid. We present these evolutions
without and with Hubble subtraction in Figure \ref{eps-n-variousLlambdas}. The
general trend is that reducing the coupling strength decreases the duration of
inflation.
\noindent The general trends in either case can be understood in terms of a simple
picture of the slow-roll approximation. Applying the slow-roll conditions to
the effective field equation (\ref{renormal0}), one obtain the speed of inflaton,
\begin{eqnarray}
\hspace{-1cm}&&|\phi'(n)|\simeq \,\Bigl | \frac{-U'(\phi)}{3\chi^2[1\!+\!g(\chi,z)]}\Bigr |
\;\;,\;\;
g(\chi,z)\!=\!h(\chi,z)\!+\!\frac13 \frac{d}{dn} h(\chi,z)\;\;,\;\;
z\!=\!\frac{\kappa}{k^3_c \chi}\,,\nonumber\\
\hspace{-1.2cm}&&h(\chi,z)\!=\!\frac{\chi^4}{4\pi^2}\frac{1}{k^3_c \chi}\!\Bigg\{
2 \gamma z + 2\Bigl[ \gamma-\zeta(3)\Bigr] z^3 +
(z \!+\! z^3) \Bigl[ \psi(1 \!+\! i z) + \psi(1 \!-\! i z)\Bigr]\Bigg\}
. \label{fermiSlowRollEom}
\end{eqnarray}
In addition to $-U'(\phi)/3\chi^2$ from a classical evolution, an
extra factor $1\!+\!g(\chi,z)$ occurs in the denominator from quantum corrections.
Because $g(\chi,z)$ is positive and $1\!+\!g(\chi,z)$ is monotonically increasing,
the quantum-corrected inflaton rolls down its potential with a smaller speed than
its classical cousin, which lengthens inflation.
The difference between the subtracted model
(the right-hand plot of Figure \ref{eps-n-variousLlambdas})
and the un-subtracted one
(the left-hand plot of Figure \ref{eps-n-variousLlambdas} )
is that the former tends to shorten inflation but with a lower peak value
for $\epsilon$. The value of $\epsilon$ falls back to zero after inflation as well.
Also note that the maximum of $\epsilon$ never reaches to $1$ until
$\lambda\leq6.98\!\times\!10^{-5} (1/k^3_c\leq3.71\!\times\!10^8)$.
\begin{figure}[H]
\includegraphics[width=4.6cm,height=4.6cm]{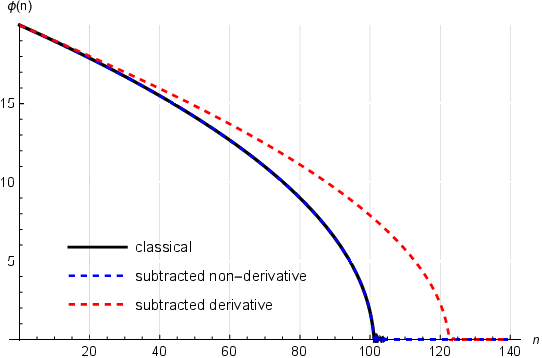}
\hspace{-0.1cm}
\includegraphics[width=4.6cm,height=4.6cm]{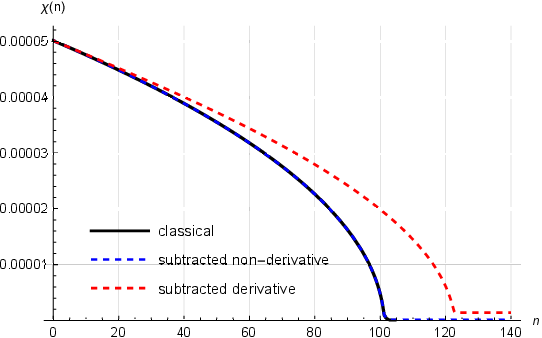}
\hspace{-0.1cm}
\includegraphics[width=4.6cm,height=4.6cm]{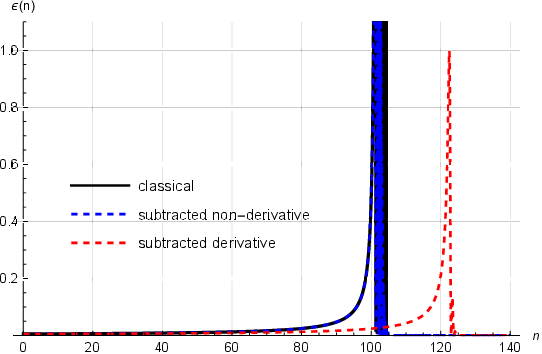}
\caption{Plots of the dimensionless scalar $\phi(n)$ (left), the
dimensionless Hubble parameter $\chi(n)$ (middle) and the first slow roll
parameter $\epsilon(n)$ (right) for a coupling constant
$\lambda \!=\! 6.98\!\times\!10^{-5}$ $(1/k^3_c\!=\!3.71\!\times\!10^{8})$ with
the initial conditions $\chi_0\!=\!5\!\times\!10^{-5}$ and
$\phi'_0\!=\!-9.99737\!\times\!10^{-2}$.}
\label{Fermiontry1}
\end{figure}
\begin{figure}[H]
\includegraphics[width=4.6cm,height=4.6cm]{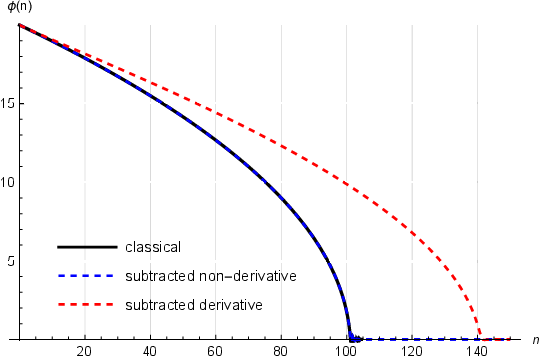}
\hspace{-0.1cm}
\includegraphics[width=4.6cm,height=4.6cm]{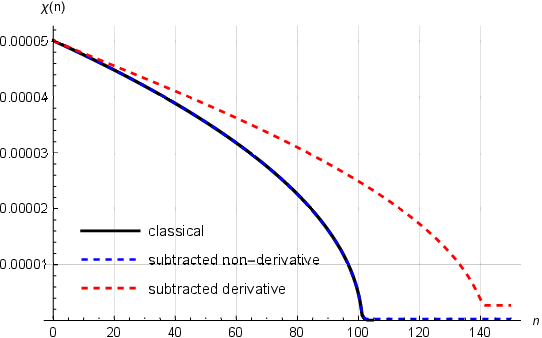}
\hspace{-0.1cm}
\includegraphics[width=4.6cm,height=4.6cm]{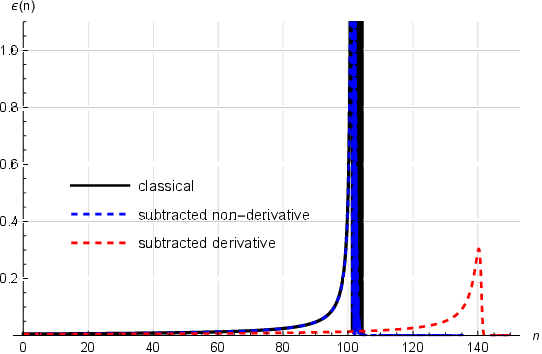}
\caption{Plots of the dimensionless scalar $\phi(n)$ (left), the
dimensionless Hubble parameter $\chi(n)$ (middle) and the first slow roll
parameter $\epsilon(n)$ (right) for a coupling constant
$\lambda\!=\! 1\!\times\!10^{-4}$ $(1/k^3_c\!=\!5.32\!\times\!10^{8})$ with
the initial conditions $\chi_0\!=\!5\!\times\!10^{-5}$ and
$\phi'_0\!=\!-9.81626\!\times\!10^{-2}$.}
\label{Fermiontry2}
\end{figure}

Finally, we compare evolution in the K-Essence model (derivative couplings)
with evolution in the analogous cosmological Coleman-Weinberg potential
(non-derivative model). These comparisons are presented in Figure \ref{Fermiontry1}
and \ref{Fermiontry2}. One can see that the derivative model deviates from a
classical evolution more than the non-derivative one at the same coupling
strength and also has a bigger dimensionless Hubble parameter after inflation.
For evolution with $\lambda \!=\! 6.98\!\times\!10^{-5}$, the dimensionless
Hubble parameter in the derivative model is $1.41\!\times\!10^{-6}$ while
it is $1.26\!\times\!10^{-7}$ in the non-derivative model. At a coupling
strength of $\lambda\!=\!1\!\times\!10^{-4}$, the dimensionless Hubble
parameters are: $\chi_{\rm derivative}=\!1.21\!\times\!10^{-5}$,
$\chi_{\rm non-derivative}=\!2.59\!\times\!10^{-7}$.
According to a previous study \cite{LMW}, the system with moderate
Yukawa coupling constant $\lambda$ tends to inflate forever. It not clear
whether we should extrapolate our current conclusion from Figure
\ref{eps-n-variousLlambdas} and draw a similar conclusion for the K-essence
model of the fermionic coupling. The obstruction to this conclusion is that,
at coupling strength $\lambda\!\sim\!0.1\Longrightarrow1/k^3_c\sim5.32
\!\times\!10^{11}$ the approximation $\chi_0\!\sim\!5\!\times\!10^{-5}$ is
not valid and the complexities of highly non-linear dependence of $\epsilon$
and $\chi$ on $\phi$ and $\phi'$ become non-trivial. No analytic solution
exists and we must employ some kind of nonlinear search routine such as a
Monte Carlo Markov chain method to solve for $\chi_0$ and $\phi'_0$ from
(\ref{F1}) and (\ref{epsilon}).

\subsection{Induced K-Essence Model due to Scalars}

The generic expression of the K-Essence from
(\ref{scalarCon}) and (\ref{scalarminimal}) takes the form,
\begin{equation}
\mathcal{K}(\kappa,\chi) = \kappa \!-\! \frac{\chi^4}{16 \pi^2}
f\Bigl(\frac{\kappa}{k^2_c \chi^2}\Bigr) \!+\! \frac{\chi^4(0)}{16\pi^2}
f\Bigl(\frac{\kappa}{k^2_c \chi^2(0)}\Bigr) \; . \label{scalarK}
\end{equation}
The two explicit cases we consider are,
\begin{eqnarray}
&&\hskip -0.5cm f_{\rm con}(y)\!=\! y\!+\!(2\gamma\!-\!1)y^2
\!+\!2\!\!\int_0^y\!\!\! dx \,x\Big[
\psi\Big(\frac12\!+\!\frac12\sqrt{1\!-\!8x}\!\Big)\!+\!
\psi\Big(\frac12\!-\!\frac12\sqrt{1\!-\!8x}\!\Big) \Big]\,,
\label{f0}\\
&&f_{\rm min}(y)\!=\! 2\gamma\!-\!\frac{182}{27}
\!+\!\Big(\!\!-4\gamma\!+\!\frac{23}{3}\Big)y
\!+\!\Big(2\gamma\!-\!\frac{25}{27}\Big)y^2 \nonumber\\
&&\hskip 1.5cm -2\!\!\int_1^y\!\!\! dx \,(1\!-\!x)\Bigg[
\psi\Bigg(\frac12\!+\!\frac32\sqrt{1\!-\!\frac89 x}\!\Bigg)\!+\!
\psi\Bigg(\frac12\!-\!\frac32\sqrt{1\!-\!\frac89 x}\!\Bigg) \Bigg],
\label{f-1}
\end{eqnarray}
where $f_{\rm con}(y)$ stands for the contribution from the conformal coupling
and the minimal coupling's contribution is denoted by $f_{\rm min}(y)$.
Similarly, by making an analogy with the non-derivative coupling term
which generates a cosmological Coleman-Weinberg potential
\begin{eqnarray}
-\frac14 h^2\varphi^2 \Phi^2 \sqrt{-g} \qquad \Longleftrightarrow \qquad
\frac{1}{2m_c^2}\partial_{\mu}\varphi\partial_{\nu}\varphi
g^{\mu\nu}\Phi^2 \sqrt{-g}\,,
\end{eqnarray}
the dimensionless couplings $h$ and $k_c$ can be related,
\begin{eqnarray}
\frac{1}{k_c^2}\sim \frac{h^2}{2k^2}\sim (1.33\!\times\!10^{10})h^2\,.
\end{eqnarray}
Also note that the scalar corrections to Coleman-Weinberg potential have
the same form as expression (\ref{scalarK}) \cite{MW1},
with the replacements of $\kappa$ by $\frac12 k^2\phi^2$
and $\frac{\kappa}{k_c^2}$ by $\frac14h^2\phi^2$,
\begin{equation}
U(\phi,\chi) = \frac12 k^2\phi^2 \!+\! \frac{\chi^4}{16 \pi^2}
f\Bigl(\frac{h^2\phi^2}{4\chi^2}\Bigr) \!-\! \frac{\chi^4(0)}{16\pi^2}
f\Bigl(\frac{h^2\phi^2}{4 \chi^2(0)}\Bigr) \; . \label{scalarU}
\end{equation}
Here the function $f$ is the same as either (\ref{f0}) or (\ref{f-1})
for conformal scalars and minimally-coupled scalars, respectively.
\begin{figure}[H]
\includegraphics[width=7.5cm,height=5.2cm]{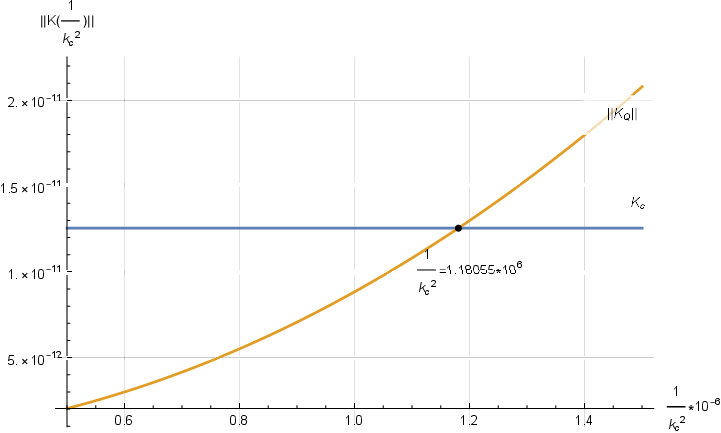}
\hspace{-0.2cm}
\includegraphics[width=7.5cm,height=5.2cm]{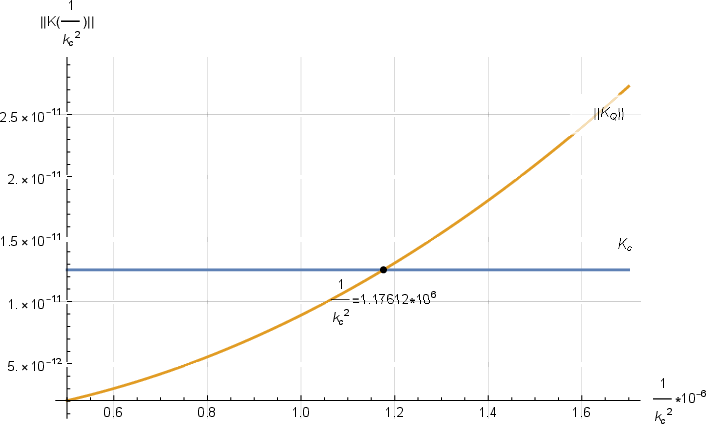}
\caption{\footnotesize Plots of $K_Q$ and $K_C$ versus $1/k^2_c$ at $n\!=\!0$
with slow-roll initial conditions (\ref{initials}). The left-hand graph is for
conformally-coupled scalars while the right-hand plot is for minimally-coupled
scalars.}
\label{abs{Kq}Kc-MiniScalar}
\end{figure}

\begin{figure}[H]
\includegraphics[width=4.6cm,height=4.6cm]{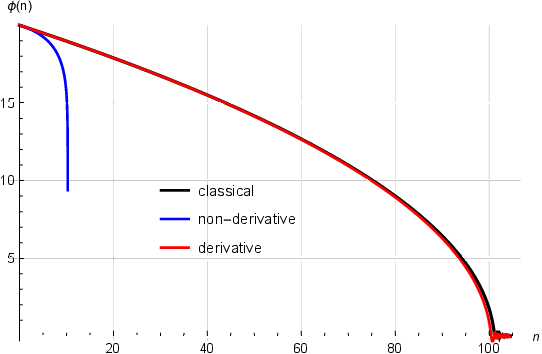}
\hspace{-0.1cm}
\includegraphics[width=4.6cm,height=4.6cm]{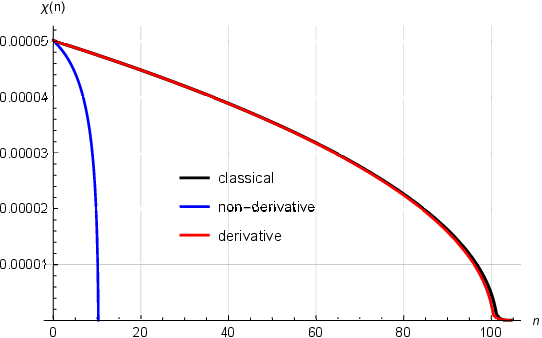}
\hspace{-0.1cm}
\includegraphics[width=4.6cm,height=4.6cm]{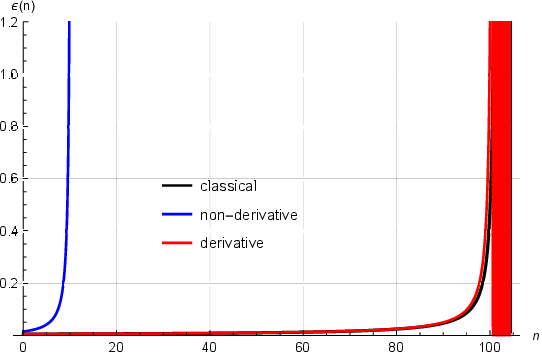}
\caption{Evolution for conformally-coupled scalars, comparing the classical
result (in black) with non-derivative coupling (in blue) and derivative
coupling (in red). Plots show the dimensionless scalar $\phi(n)$ 
(left), the dimensionless Hubble parameter $\chi(n)$ (middle) and the first
slow roll parameter $\epsilon(n)$ (right) for an effective coupling
constant of $h^2\!=\! 1.6\!\times\!10^{-5}$, corresponding to $1/k^2_c\!=
\!2.218\! \times\!10^{5}$, with slow-roll initial conditions
(\ref{initials}).}
\label{ConScalarTry}
\end{figure}

\begin{figure}[H]
\includegraphics[width=4.6cm,height=4.6cm]{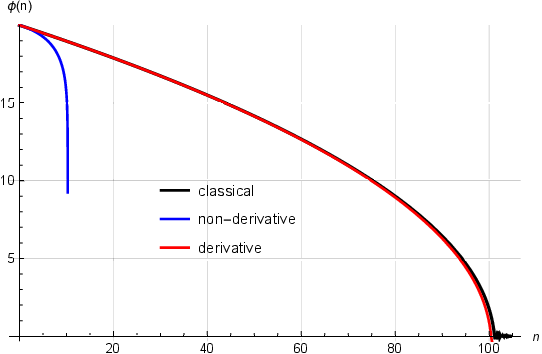}
\hspace{-0.1cm}
\includegraphics[width=4.6cm,height=4.6cm]{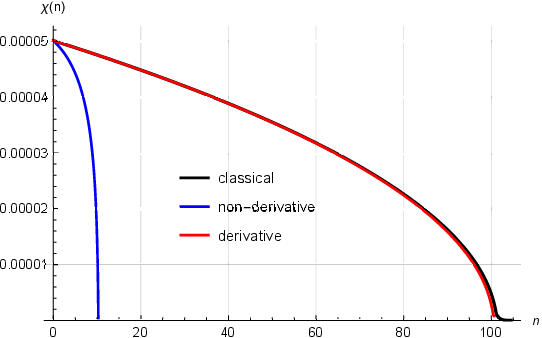}
\hspace{-0.1cm}
\includegraphics[width=4.6cm,height=4.6cm]{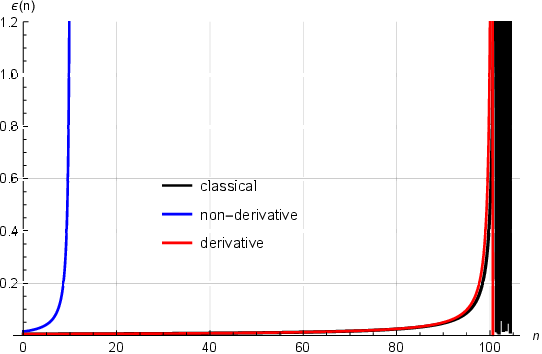}
\caption{Evolution for minimally-coupled scalars, comparing the classical
model (in black) with non-derivative couplings (in blue) and derivative
coupling (in red). Plots show the dimensionless scalar $\phi(n)$ (left),
 the dimensionless Hubble parameter $\chi(n)$ (middle) and the
first slow roll parameter $\epsilon(n)$ (right) for an effective
coupling constant of $h^2\!=\! 1.6\!\times\!10^{-5}$, corresponding to
$1/k^2_c\!=\!2.218 \!\times\!10^{5}$, with slow-roll initial conditions
(\ref{initials}).}
\label{MiniScalarTry}
\end{figure}

We first determine when evolution enters the quantum-dominated domain for
conformal scalars and minimally-coupled scalars. Figure \ref{abs{Kq}Kc-MiniScalar}
shows that this happens around $1/k^2_c\!\gtrsim\!10^6$ $(h^2\!\gtrsim\!10^{-5})$
in both cases. Because $K_Q$ in the scalar K-Essence model is negative we must
not go beyond the classical-dominated regime in order to avoid a disastrous
kinetic instability. When $K_C$ dominates over $K_Q$, it is reliable to use the
slow-roll initial conditions (\ref{initials}). Within the safe regime, we find
that quantum effects due to each sort of scalar tend to shorten inflation.
The result agrees with what one expects from the effective field equation using
the slow roll approximation,
\begin{eqnarray}
\mu(\chi,z)\phi'(n)\!\sim\!\frac{-U'(\phi)}{3\chi^2}\;\;\;,\;\;\;
\mu(\chi,z)\!\equiv\! 1\!+\!\alpha(\chi,z)\;\;\;,\;\;\;
z\!=\!\frac{\kappa}{k^2_c \chi^2}\,.
\end{eqnarray}
\noindent Because quantum contributions are negative, they make the inertia
$\mu(\chi,x)$ smaller than $1$ and hence cause the inflaton to roll down
its potential more rapidly than that in the classical model.
Figure \ref{ConScalarTry} compares the K-Essence model (derivative), with
the model of non-derivative couplings (non-derivative) and the
classical result for an inflaton coupled to conformal scalars whose
coupling strength is $h^2\!=\! 1.6\!\times\!10^{-5}$
$(1/k^2_c\!=\!2.218\!\times\!10^{5})$. Figure \ref{MiniScalarTry} presents
a similar comparison, at the same coupling constant, for an inflaton coupled to
minimally-coupled scalars. There is little distinction between conformal and
minimal coupling. At this coupling strength inflation ends at about 10
e-foldings for the non-derivative model, whereas the derivative model nicely traces
the classical evolution.

\section{Conclusions}

Quantum-induced K-Essence models, or cosmological Coleman-Weinberg potentials,
are the price scalar-driven inflation pays for efficient reheating.
Although coupling the differentiated inflaton to matter alters the kinetic energy
instead of the potential, quantum-induced K-Essence and cosmological
Coleman-Weinberg potentials are the same functions of different arguments.
On de Sitter this goes like $H^4$
multiplying a complicated function of the dimensionless parameters,
$\textsc{k}/m^3_c H$ for fermions and $\textsc{k}/m^2_c H^2$ for scalars.
Recent work \cite{Kyriazis:2019xgj,Sivasankaran:2020dzp,Katuwal:2021kry}
strongly supports the idea that these de Sitter results remain approximately valid
when the constant Hubble parameter of de Sitter is replaced by the evolving $H(t)$
of realistic inflation. A significant difference between quantum-induced K-Essence
models and cosmological Coleman-Weinberg potentials is that the dimensionless
parameters of the former {\it increase} as time progresses, so that the system
remains in the large field domain over the course of inflation. As a result,
changes due to the quantum correction are enormously large.

Because $H(t)$ is not a local functional of the metric for a general geometry,
neither quantum-induced K-Essence nor cosmological Coleman-Weinberg potentials can
be completely eliminated using local counterterms \cite{MW1}. The technique of
{\bf Hubble subtraction} consists of subtracting the local counterterm which
results from setting $H(t)$ to its initial value. One can compare the
dimensionful coupling constants of quantum-induced K-Essence with the dimensionless
coupling constants of cosmological Coleman-Weinberg potentials by requiring the
two couplings to have the same strength during reheating. This paper is devoted to
comparing quantum-induced K-Essence with cosmological Coleman-Weinberg potentials,
with and without Hubble subtraction, regarding the competing requirements of
facilitating efficient reheating without disturbing the observational constraints
(\ref{duration}-\ref{tensortoscalar}) on primordial inflation.

Fermionic K-Essence (derivative) and fermionic Coleman-Weinberg potentials
(non-derivative) both tend to lengthen inflation, albeit for different reasons.
One can understand this by looking at the scalar evolution equation simplified by
the slow roll approximation,
\begin{eqnarray}
\phi'(n)\approx\frac{-1}{3\chi^2}
\frac{\Big[U'_{\rm cl}(\phi)\!+\!\Delta U'(\phi)\Big]}
{1 \!+\! g(\chi,z)} \qquad , \qquad z \equiv  \frac{\kappa}{k^3_c \chi}
\Big(\frac{\textsc{k}}{m^3_c H}\Big) \,.
\label{genericScalarEOM}
\end{eqnarray}
Non-derivative couplings induce $\Delta U'(\phi)<0$ and $g(\chi,z)\!=\!0$, so they
decrease $|\phi'(n)|$ (and hence increase the duration of inflation) by decreasing
the force term. In contrast, derivative couplings induce $\Delta U'(\phi)\!=\!0$
with $g(\chi,z)\!>\!0$, so $|\phi'(n)|$ is also decreased, but now by increasing the
inflaton's inertia $\mu(\chi,z)=1\!+\!g(\chi,z)$. However, fermionic K-Essence
enters the quantum-dominated regime even at a quite small effective coupling
$\lambda\!\sim\!6.3\!\times\!10^{-5}$, for which the cosmological Coleman-Weinberg
potential would be small compared to the classical potential. At the same coupling
strength, the quantum effect in the derivative model is generically stronger than
it is for the non-derivative one. This is why in Figures
\ref{Fermiontry1} and \ref{Fermiontry2} show longer durations for inflation,
and larger dimensionless Hubble parameters after the end of inflation, for fermionic
K-Essence models than for the analogous cosmological Coleman-Weinberg potentials.
For non-derivative couplings inflation does not even end ($\epsilon = 1$) unless
the coupling is less than about $\lambda\!\sim\!1.15\!\times\!10^{-4}$ \cite{LMW},
whereas this occurs for derivative couplings at an even smaller equivalent value of
$\lambda\!\sim\!6.98\!\times\!10^{-5}$. Neither fermionic model (derivative or
non-derivative) begins to have a reasonable evolution until the coupling is made
very small, which could endanger reheating.

For scalars, both derivative and non-derivative couplings induce models which are
within the classical domain at the small effective coupling constant of
$h^2=1.6\!\times\!10^{-5}$. Scalar-induced K-Essence suffers from a disastrous
kinetic instability (that is, $g(\chi,z) < -1$ in equation (\ref{genericScalarEOM})
with $z = \textsc{k}/m^2_c H^2$) if it leaves the classical-dominated regime. From
Figures \ref{ConScalarTry} and \ref{MiniScalarTry}, one can see that both
derivative and non-derivative models tend to shorten inflation, with an especially
rapid termination of inflation for the non-derivative model. One can understand
this from the analog of equation (\ref{genericScalarEOM}) with $z = \textsc{k}/m^2_c
H^2$. Non-derivative couplings induce $\Delta U > 0$ with $g = 0$, so they
increase $|\phi'(n)|$ by strengthening the force. Derivative couplings induce
$\Delta U = 0$ with $g < 0$, which increases $|\phi'(n)|$ by reducing the inertia.
We found no significant differences between minimally-coupled and conformally
coupled scalars, presumably because the (derivative or non-derivative) coupling
is larger than the conformal coupling. Generally speaking, coupling the
differentiated inflaton to scalars seems to do a better job than the non-derivative
coupling. However, neither coupling provides a very
satisfactory resolution of the
tension between facilitating efficient reheating and preserving the observational
constraints (\ref{duration}-\ref{tensortoscalar}).

Because dimensionful coupling constants are introduced to quantum-induced K-Essence,
we would like to digress in order to comment on the validity of these models as low
energy effective field theories.
Even though our couplings (\ref{fermionL}) and (\ref{phi}) are not renormalizable,
they induce no higher derivative counterterms (\ref{f2ct}) and (\ref{2ct}) as long as
the inflaton kinetic factor (\ref{mphi}) is considered to be constant. Without
inflaton loops, the only way one gets higher loop corrections is from the interaction
with gravity. Hence the 2-loop contribution would take the form of
$G\Big(\frac{\partial\varphi\cdot\partial\varphi}{m_c^3}\Big)^6$ and its effect
is suppressed by $G$. As a result, we recognize the cutoff of this effective field theory
as the Planck mass. Inflation is assumed to occur well below the Planck mass.
Note that the value of the quantum gravitational loop counting parameter $GH^2$
(obtained from the upper bound of the tensor to scalar ratio and the scalar power
spectrum \cite{Planck:2018vyg})
is about $10^{-11}$. So the work considered here should be safely within the realm of validity
of low energy effective field theory.

There is so far no consensus about the magnitude of the reheat temperature.
The common belief is that it could be as low as $10^2$ GeV and as high as
$10^{19}$ GeV. If one extrapolates the initial form of the potential to
the end of inflation, one finds a large reheat temperature. This is done
by employing observational data on primordial perturbations to estimate the
number of e-foldings since the end of inflation, and then comparing that
with a thermal estimate of the same quantity. Using WMAP data, Martin and
Ringeval derived a bound of $T_R > 10^4~{\rm GeV}$ \cite{Martin:2010kz},
and more recent data raise this bound. Please see the Appendix for a
detailed explanation. Accommodating a lower reheat temperature requires
dramatic changes in the shape of the potential after the emission of
currently observable perturbations.

Leaving aside the issue of what geometric and thermal considerations imply for
the value of the reheat temperature, we can estimate $T_R$ from the dynamics
of the coupling. For the fermionic K-Essence model considered in this paper,
if one imagines that the interaction is a sort of effective Yukawa coupling,
the reheat temperature can be estimated when the Hubble parameter falls below
the decay rate $\Gamma\!=\!\frac{\lambda^2 m}{8\pi^2}$ 
(where $m$ is the mass of the inflaton)
\cite{LMW,Kofman:1997yn,Greene:1997fu},
\begin{equation}
T_R \simeq \frac15 \Bigl( \frac{\Gamma^2}{G}\Bigr)^{\frac14} \simeq \lambda
\times 10^{15}~{\rm GeV} \; . \label{Treheat}
\end{equation}
As we have seen, the largest value of $\lambda$ which is consistent with viable
inflation is $\lambda\sim 10^{-5}$. This corresponds to reheat temperature
of $T_R\sim 10^{10}$ GeV.

An alternative approach would be to make no subtractions and attempt instead to
cancel the positive kinetic corrections induced by fermions with the negative
ones induced by scalars. Of course coupling the inflaton to more fields aids
reheating, so there are no worries on that score. By adjusting the effective
coupling constants $\lambda$ and $h^2$'s, one could manage to get the leading
terms of the large field expansions (\ref{Klargef}), (\ref{ConL}) and (\ref{minL})
to cancel. The question then becomes how the lower order terms affect the
observational constraints (\ref{duration}-\ref{tensortoscalar}).

In addition to studies on axionic couplings like
$\varphi\epsilon^{\mu\nu\rho\lambda}F_{\mu\nu}F_{\rho\sigma}$ by
Adshead and collaborators \cite{Adshead},
another alternative is to investigate derivative couplings to vector bosons,
analogous to the undifferentiated couplings already studied for a charged
inflaton \cite{MW1, LMW, MPW, Katuwal:2021kry, MTanW},
\begin{eqnarray}
\mathcal{L} = \frac{R \sqrt{-g}}{16 \pi G} \!-\! \frac14 F_{\rho\sigma}
F_{\mu\nu} g^{\rho\mu} g^{\sigma\nu} \sqrt{-g}
- \Bigl(\partial_{\mu} \!-\! i q A_{\mu} \Bigr) \varphi^*
\Bigl( \partial_{\nu} \!+\! i q A_{\nu}\Bigr) \varphi g^{\mu\nu} \sqrt{-g}\,.
\label{GaugeBoson}
\end{eqnarray}
One might explore some exotic derivative coupling of uncharged inflatons such as,
\begin{eqnarray}
-\frac{1}{4m^4_c}\partial_{\alpha}\varphi\partial_{\beta}\varphi g^{\alpha\beta}
\times F_{\rho\sigma}F_{\mu\nu} g^{\rho\mu} g^{\sigma\nu} \sqrt{-g}\,.
\end{eqnarray}
We might consider its undifferentiated cousin to be,
\begin{eqnarray}
-\frac{1}{4m^2_c}\varphi^2
F_{\rho\sigma}F_{\mu\nu} g^{\rho\mu} g^{\sigma\nu} \sqrt{-g}\,.
\end{eqnarray}
To carry out studies with such derivative couplings, one first needs to figure out
the coincidence limit of the field strength propagator on de Sitter,
\begin{eqnarray}
\Big<F_{\rho\sigma}(x)F_{\mu\nu}(x) g^{\rho\mu} g^{\sigma\nu}\Big>\,.
\end{eqnarray}

\vskip 1cm

\centerline{\bf Acknowledgements}

This work was supported by by Taiwan MOST grants 110-2112-M-006-026
and NSTC 111-2112-M-006-038.

\section*{Appendix: Connecting Data to the Reheat Temperature $T_R$ \label{Appendix}}

In this appendix, we begin by presenting two approaches to estimate the number
of e-foldings from the end of inflation to the current time. Comparison of these
results implies that a large $T_R$ is favored. To facilitate the discussion we
define $N \equiv \ln[\frac{a(t)}{a_i}]$ as the number of e-foldings since the
start of inflation to co-moving time $t$. If we follow the usual practice that
the current scale factor is one ($a_0 = 1$) then the number of e-foldings between
any event and now is,
\begin{equation}
N_0 - N = \ln\Bigl[\frac{1}{a(t)}\Bigr] \; .
\end{equation}

We begin by using the geometry of inflation to estimate the number of e-foldings
from the end of inflation to now \cite{Mielczarek:2010ag}. Primordial perturbations
which are today observed with co-moving wave number $k$ experienced first horizon
crossing at a time $t_k$ during inflation with $k\!\equiv\!H(t_k)a(t_k)$. The
number of e-foldings from then to now is,
\begin{eqnarray}
N_0 - N_k = \ln\Big[\frac{H(t_k)}{k}\Big] \simeq \frac12
\ln\Big[\Delta^2_{\mathcal{R}}(k)\frac{\pi\epsilon(t_k)}{Gk^2}\Big]\,,
\end{eqnarray}
where the final step follows from the approximate form of the scalar power
spectrum $\Delta^2_{\mathcal{R}}(k)\cong\frac{GH^2(t_k)}{\pi\epsilon(t_k)}$.
Now substitute the simple power law formula used to represent the observational
data in terms of a scalar amplitude $A_s \simeq 2 \times 10^{-9}$ and spectral
index $n_s \simeq 1 - 0.035$ around a pivot wave number of $k_0 =
0.05~{\rm Mpc}^{-1}$,
\begin{eqnarray}
\Delta^2_{\mathcal{R}}(k)\cong A_s\Big(\frac{k}{k_0}\Big)^{n_s-1} \Longrightarrow
N_0 - N_{k_0} = \frac12
\ln\Big[\frac{A_s \pi r}{16 G k_0^2}\Big] \simeq 122.33 +
\frac12 \ln\Bigl[\frac{r}{0.036}\Bigr] \,.\label{n1}
\end{eqnarray}
The final logarithm would vanish if the tensor-to-scalar ratio were resolved at its
current upper limit; otherwise it would be negative. We stress that expression
(\ref{n1}) derives from geometry and observation; it cannot be evaded (within the
context of single-scalar inflation), no matter what assumptions are made about the
inflationary potential or the mechanism of reheating. It is rather our assumptions
about these two things which must accommodate expression (\ref{n1}).

Consider first the duration of inflation after the pivot wave number experienced
first horizon crossing. We will see that increasing this interval
increases the reheat temperature.
Trivial calculus allows us to express this as an integral
over the first slow roll parameter,
\begin{eqnarray}
N_{e} - N_{k_0} = \int_{\epsilon(t_{k_0})}^1 \frac{d\epsilon}{\epsilon'} \, .
\label{nA}
\end{eqnarray}
Primes on $\epsilon$ denotes derivatives with respect to the number
of e-foldings.
At this point the inflationary potential becomes relevant. When these estimates
were first made in 2010 \cite{Mielczarek:2010ag}, the quadratic potential was
employed to conclude $\epsilon' = 2 \epsilon^2$, and the resulting integral was
evaluated to $2/(1 - n_s) -\frac12 \cong 56.64$.
Since then, the increasingly tight
upper bounds on $r$ (and hence on $\epsilon = r/16$) have favored very flat
potentials which give an even larger result. To understand this, consider the
slow roll relation for the spectral index under the assumption of very small
$\epsilon$ with $n_s$ fixed (which is supported by the absence of evidence
for running of the spectral index \cite{Planck:2018vyg}),
\begin{eqnarray}
1 \!-\! n_s = 2 \epsilon + \frac{\epsilon'}{\epsilon} \qquad \Longrightarrow
\qquad \epsilon' = (1 \!-\! n_s) \epsilon - 2 \epsilon^2 \simeq (1 \!-\! n_s)
\epsilon \, . \label{nB}
\end{eqnarray}
Substituting this approximation in the integral (\ref{nA}) gives,
\begin{eqnarray}
N_{e} - N_{k_0} \simeq \frac1{1 \!-\! n_s} \int_{\epsilon(t_{k_0})}^1
\frac{d\epsilon}{\epsilon} = \frac{ \ln(\frac{16}{r})}{1 \!-\! n_s} \simeq
28.57 \Biggl( 6.10 - \ln\Bigl[ \frac{r}{0.036}\Bigr]\Biggr) \, . \label{n2}
\end{eqnarray}
Of course expression (\ref{n2}) is larger than (\ref{n1}),
which is impossible, but it indicates the general trend
towards large reheat temperatures. To facilitate the discussion we made a
numerical computation using the Einstein frame version of Starobinsky's
model \cite{Starobinsky} which gives $N_e - N_{k_0}\cong 51.32$.
Because this number is smaller than either the result
for quadratical potential or our general estimate (\ref{n2}), we can regard
it as a sort of lower bound.
Combining this figure with (\ref{n1}) gives
\begin{eqnarray}
\Delta N \equiv N_0 - N_{e} = \frac12
\ln\Big[\frac{A_s \pi r}{16 G k_0^2}\Big] - 51.32
\simeq 71.01 +
\frac12 \ln\Bigl[\frac{r}{0.036}\Bigr] \,.\label{DeltaN1}
\end{eqnarray}
Note that inserting the much lower value of $r$ predicted
for the Starobinsky's model would decrease $\Delta N$, thereby increasing
the reheat temperature.

We can use thermodynamics to estimate the number of e-foldings from the
end of inflation to now $\Delta N \equiv N_0 - N_e$. The estimate is based on
considering three periods \cite{Mielczarek:2010ag}:
\begin{enumerate}
\item{The interval from the end of inflation to the end of reheating;}
\item{The interval from the end of reheating to recombination; and}
\item{The interval from recombination to now.}
\end{enumerate}
The (approximately matter dominated) energy density at the end of inflation
$\rho_e$ derives from the kinetic energy of the inflaton. We assume that
this kinetic energy is approximately constant throughout inflation, which
implies that it can be related to conditions at the time the pivot wave
number experiences first horizon crossing,
\begin{eqnarray}
\rho_e = \frac12 \dot{\varphi}^2 = \frac{\epsilon H^2}{8\pi G} \simeq
\frac{r^2 A_s}{2^{11} G^2} \, . \label{rhoEnd}
\end{eqnarray}
This energy density redshifts like nonrelativistic matter until reheating
converts it into the energy density of $g_*$ relativistic species,
\begin{eqnarray}
\rho_R  = \rho_e \Bigl(\frac{a_e}{a_R}\Bigr)^3 = \frac{g_{*}\pi^2T_R^4}{30}
\,,\label{rhoR}
\end{eqnarray}
Hence the number of e-foldings from the end of inflation to the end of
reheating is,
\begin{eqnarray}
N_R - N_e = \frac13 \ln\Bigg[\frac{15 r^2 A_s}{2^{10} \pi^2 g_{*}G^2 T_R^4}\Bigg]
\, . \label{na}
\end{eqnarray}
Entropy is conserved during the second period, which means,
\begin{eqnarray}
\frac{a_{\rm rec}}{a_R} = \Big(\frac{g_*}{2}\Big)^{\frac13}
\times \frac{T_R}{T_{\rm rec}} \;\; \Longrightarrow \;\;
N_{\rm rec} - N_R = \frac13
\ln\Bigg[\frac{g_* T_R^3}{2 T^3_{\rm rec}}\Bigg] \, . \label{nb}
\end{eqnarray}
Finally, from recombination to now we have,
\begin{eqnarray}
\frac{a_0}{a_{\rm rec}} = \frac{T_{\rm rec}}{T_0} \;\; \Longrightarrow \;\;
N_0 - N_{\rm rec} = \frac13 \ln\Bigg[\frac{T^3_{\rm rec}}{T_0^3}\Bigg]
\, . \label{nc}
\end{eqnarray}
Adding the three intervals (\ref{na}), (\ref{nb}) and (\ref{nc}) gives,
\begin{equation}
\Delta N \equiv N_0 - N_e = \frac13 \ln\Biggl[ \frac{15 r^2 A_s}{2^{11}
\pi^2 G^2 T_{R} T_0^3}\Biggr] \simeq 62.65 + \frac23 \ln\Bigl[\frac{r}{0.036}\Bigr]
- \frac16 \ln(G T^2_R) \; , \label{DeltaN2}
\end{equation}
Equating (\ref{DeltaN1}) and (\ref{DeltaN2}), and assuming that the tensor
power spectrum is resolved at the current upper limit, we find the reheat
temperature to be $T_R\sim 10^{8}~{\rm GeV}$.
One can avoid this conclusion by fine tuning the model to reduce $N_e - N_{k_0}$,
but the preference for a large reheat temperature is clear.

\end{document}